\documentclass[a4paper,fleqn,usenatbib]{mnras}

\usepackage{newtxtext,newtxmath}

\usepackage[T1]{fontenc}
\usepackage{ae,aecompl}

\usepackage{graphicx}	
\usepackage{amsmath}	
\usepackage{natbib}

\defcitealias{Belloni2017a}{Belloni et~al.}
\defcitealias{Belloni2017b}{Belloni et~al.}
\newcommand\citetBab[2]{\citeauthor{#1}\ (\citeyear{#1}a, b)}

\newcommand\citetb[1]{\citetalias{#1}\ (\citeyear{#1}b)}

\newcommand\citealtb[1]{\citetalias{#1}\ \citeyear{#1}b}

\usepackage{xspace}

\usepackage[export]{adjustbox}

\newcommand{\CONCPOP}{$\text{conc}_{\text{pop}}$\xspace}

\newcommand{\MOCCA}{\textsc{mocca}\xspace}
\newcommand{\NBODY}{\textsc{nbody}\xspace}
\newcommand{\SURVEYONE}{\textsc{mocca-survey-1}\xspace}
\newcommand{\SURVEYTWO}{\textsc{mocca-survey-2}\xspace}
\newcommand{\RHOB}{$\rm R_{hob}$\xspace}

\title[MOCCA -- Multiple stellar populations
]{MOCCA: Dynamics and evolution of single and binary stars of multiple stellar populations in tidally filling and underfilling globular star clusters}

\author[Arkadiusz Hypki]{Arkadiusz Hypki$^{1}$\thanks{E-mail: ahypki@camk.edu.pl},
Mirek Giersz$^{1}$,
Jongsuk Hong$^{2}$, Agostino Leveque$^{1}$,
\newauthor
Abbas Askar$^{3}$, Diogo Belloni$^{4}$ and Magdalena Otulakowska-Hypka$^{5}$\\
\\
$^{1}$Nicolaus Copernicus Astronomical Center, Polish Academy of Sciences, Bartycka 18, 00-716 Warsaw, Poland\\
$^{2}$Korea Astronomy and Space Science Institute, Daejeon 34055, Republic of Korea\\
$^{3}$Lund Observatory, Department of Astronomy, and Theoretical Physics, Lund University, Box
43, SE-221 00 Lund, Sweden\\
$^{4}$National Institute for Space Research, Av. dos Astronautas, 1758, 12227-010, São José dos Campos, SP, Brazil\\
$^{5}$Astronomical Observatory Institute, Faculty of Physics, A. Mickiewicz
University, S\l{}oneczna 36, 60-286 Pozna\'n, Poland
}

\date{Accepted XXX. Received YYY; in original form ZZZ}

\pubyear{2022}

\begin{document}
\label{firstpage}
\pagerange{\pageref{firstpage}--\pageref{lastpage}}
\maketitle

\begin{abstract}

We present an upgraded version of the \MOCCA code for the study of dynamical evolution of globular clusters (GCs) and its first application to the study of evolution of multiple stellar populations. We explore initial conditions spanning different structural parameters for the first (FG) and second generation of stars (SG) and we analyze their effect on the binary dynamics and survival. Here, we focus on the number ratio of FG and SG binaries, its spatial variation, and the way their abundances are affected by various cluster initial properties. We find that present-day SG stars are more abundant in clusters that were initially tidally filling. Conversely, FG stars stay more abundant in clusters that were initially tidally underfilling. We find that the ratio between binary fractions is not affected by the way we calculate these fractions (e.g. only main-sequence binaries (MS) or observational binaries, i.e. MS stars $> 0.4 M_{\odot}$ mass ratios $> 0.5$). This implies that the MS stars themselves are a very good proxy for probing entire populations of FG and SG. We also discuss how it relates to the observations of Milky Way GCs. We show that \MOCCA models are able to reproduce the observed range of SG fractions for Milky Way GCs for which we know these fractions. We show how the SG fractions depend on the initial conditions and provide some constraints for the initial conditions to have more numerous FG or SG stars at the Hubble time.

\end{abstract}

\begin{keywords}
stellar dynamics - methods: numerical - globular clusters: evolution - stars:
multiple stellar populations
\end{keywords}



\section{Introduction}
\label{sec:Intro}

Globular clusters (GCs) are old and dense stellar systems that play an important role in the study of stellar evolution, stellar dynamics and stellar populations resulting from complex dynamical interactions. They are very efficient laboratories to study all types of stellar astronomical objects, even the most exotic ones, like cataclysmic variables, X-ray binaries, blue straggler stars (BSSs), black holes (BHs) or intermediate-mass black holes (IMBHs).

Initially, GCs were considered to host only a single stellar population which had to be formed from a single cloud upon the cluster formation. This implied that they should have the same chemical composition and age for all the stars in the cluster. It was indeed observed in the first color-magnitude diagrams (CMDs) of GCs that they showed a common picture: the main sequence (MS), sub-giant, red giant, and horizontal branches appeared as single lines. 
The first direct observational proofs of multiple stellar populations (MSPs) were given by \citet{Lee1999Natur.402...55L} for $\omega~\rm Cen$.
Extensive observational study of HST results on GCs abundances variations and MSP was presented by \citet{2015AJ....149...91P} for many GCs (see also \citealt{Milone2017MNRAS.464.3636M}). In turn, \citet{Renzini2015} compared the possible formation scenarios of MSPs with regards to the new observational facts. Now, it appears that nearly all galactic GCs show evidence of MSPs, e.g. by multiple MS populations in their CMDs, chemical variations, anti-correlations of elements or extreme helium abundance (see \citet{Bastian2013,Bastian2018,Gratton2019A&ARv..27....8G} and references therein). Sometimes, there are more than two populations present, e.g. NGC~2808 with at least five populations \citep{Milone2015,2016MNRAS.463..449S} or M2 with seven \citep{2015MNRAS.447..927M}. MSPs are also found in GCs outside of our Galaxy, e.g. in NGC~1846, 1806 and 1783 in Large Magellanic Cloud (LMC) \citep{Mackey2008ApJ...681L..17M}, in LMC intermediate age clusters \citep{Milone2009A&A...497..755M,2015MNRAS.450.3750M}, or in M31 \citep{2013A&A...549A..60C}.

There are a number of observational techniques which allow to identify the multiple populations. Thanks to HST photometry of 59 GCs it was revealed that first (FG), and second generation of stars (SG) appear as distinct sequences in many regions of the CMDs \citep{Milone2017MNRAS.464.3636M}. Together with photometric observations also the spectroscopic methods were used to study MSP revealing differences in the abundances of p-capture elements \citep{Gratton2004MmSAI..75..274G,Carretta2009A&A...505..117C,Gratton2012A&ARv..20...50G,Gratton2019A&ARv..27....8G}. All of this demonstrates that MSPs are rather common in Galactic GCs. The number of multiple populations for GCs varies quite substantially -- from two populations for low mass clusters up to several populations in massive clusters (e.g. $\omega~\rm Cen$). However, there are studies showing that there might be some GCs with no apparent signs of MSPs, which adds some complexity to this picture, e.g. Ruprecht~106 \citep{Villanova2013ApJ...778..186V}. 

Spatial and kinematic differences have been observed in several GCs \citep[e.g.][]{Richer2013ApJ...771L..15R,Bellini2015ApJ...810L..13B,Cordero2017MNRAS.465.3515C,2018MNRAS.479.5005M,Dalessandro2021MNRAS.506..813D} while in some no differences were found \citep[e.g][]{Cordoni2020ApJ...889...18C}. The spatial differences in velocities are consistent with formation models proposed by \citep[e.g.][and references therein]{Calura2019MNRAS.489.3269C}. A number of studies have investigated the formation \citep{Decressin2007,DErcole2008MNRAS.391..825D,Bekki2010ApJ...724L..99B,Bekki2011MNRAS.412.2241B,Gieles2018,Calura2019MNRAS.489.3269C,Wang2020MNRAS.491..440W,McKenzie2021MNRAS.507..834M} and different aspects of the dynamical evolution of multiple stellar populations \citep{Vesperini2013MNRAS.429.1913V, Vesperini2018MNRAS.476.2731V, Vesperini2021MNRAS.502.4290V, HenaultBrunet2015MNRAS.450.1164H, Miholics2015MNRAS.454.2166M, Tiongco2019, Sollima2021MNRAS.502.1974S}. The effects of the tidal field on the formation of MSP were investigated e.g. by \citet{Vesperini2021MNRAS.502.4290V,Sollima2021MNRAS.502.1974S}.

Some of the properties of MSPs among GCs have a wide spread and can differ strongly between clusters. For example the fraction of SG stars have values as low as 35\% for some GCs (e.g. M~71) and as high as 90\% for others ($\omega~\rm Cen$) which poses a great challenge for interpretation. Also the radial distributions of FG, and SG presents  some interesting properties. There are GCs which have SG more radially concentrated than the FG, e.g. $\omega~\rm Cen$ \citep{Sollima2007ApJ...654..915S}, 47~Tuc \citep{Cordero2014ApJ...780...94C}, and there are GCs for which radial profiles of the two populations suggest no differences, e.g. NGC~6752  \citep{Nardiello2015A&A...573A..70N}, M~5 \citep{Lee2017ApJ...844...77L}, or NGC~6362 \citep{Dalessandro2014}. 

There is no significant correlation between GCs masses and the galaxy orbital parameters \citep{Milone2017MNRAS.469..800M}. However, there is some sign that larger fraction of FG stars is characteristic for the GCs with larger perigalactic radii \citep{Zennaro2019MNRAS.487.3239Z,Milone2020MNRAS.491..515M}. The possible role of mass loss due to stellar escape from two-body relaxation in producing this trend is discussed also in \citet{Vesperini2021MNRAS.502.4290V}. Strong correlation between fraction of SG with the cluster total mass was reported by \citet{Carretta2010A&A...516A..55C} and \citet{Milone2017MNRAS.464.3636M}. This suggests that the host galaxies have some kind of influence on the GCs evolution after all. \citet{Baumgardt2018MNRAS.478.1520B} reports also that simple-population GCs seem to have initial masses smaller than $\sim 1.5 \cdot 10^5 M_{\odot}$, whereas MSPs are present in more massive ones. However, this conclusion is also challenged by e.g. NGC~419 \citep{Li2020ApJ...893...17L} which is a massive cluster but with no apparent signs of MSPs.

We can trace formation and dynamical history of stars with the use of numerical simulations, and their physical nature with various astronomical observations. Star clusters are often used to test stellar evolution and population synthesis models as well as MSPs within the GCs. The aspect of MSPs and their formation is especially important in the context of our understanding of the formation and evolution of GCs and their host galaxies. 


Binaries play an important role in the dynamical evolution of globular clusters, and they are also one of the important source of the production of exotic astronomical objects. Since it is very difficult to determine their populations and binarity together in the photometric observations, the nature and dynamics of binary stars in the multiple stellar populations has been relying on the theoretical approaches \citep{Vesperini2011MNRAS.416..355V,Hong2015MNRAS.449..629H,Hong2016MNRAS.457.4507H,Hong2019MNRAS.483.2592H,Khalaj2015MNRAS.452..924K}. Especially, \citet{Sollima2022} investigated with Monte Carlo codes the fractions of SG binaries surviving to the Hubble time. They claim that the fraction of SG binaries depends only on the ratio between the total cluster mass and the initial size of the SG. A remarkable attempt to distinguish the multiple population binaries has been made by \citet{Milone2020MNRAS.492.5457M}. They obtained the distribution of populations among the binaries using {\it chromosome maps} and proved the existence of mixed binaries composed of one FG and one SG stars. Another interesting observational study for the binaries in the multiple populations was done by \citet{Dalessandro2018ApJ...864...33D}. They found an inflation of FG/SG ratio of velocity dispersion at the large radii in the fully-mixed cluster NGC 6362 whose velocity dispersions are supposed to be identical for the different population. It turned out, that the inflation is due to the internal motion of binaries with the fact that the FG binaries are more abundant at the large radii compared to the SG binaries which are preferentially disrupted at the central regions \citep{Hong2019MNRAS.483.2592H,MastrobuonoBattisti2021}.

There are proposed multiple explanations how the SG could be formed from the leftover gas from the FG formation and enriched: by AGB stars \citep{Ventura2001}, fast rotating massive stars and massive binaries \citep{Decressin2007,Krause2013}, supermassive stars  \citep{Denissenkov2014,Gieles2018}, accretion disks  \citep{Bastian2013}, or non-conservative mass-transfer in massive binaries \citep{deMink2009}. There is also a scenario which assumes that star clusters could form as a merger of protoclusters \citep{Elmegreen2017,Howard2019}. However, there is no conclusive evidence which of those scenarios are responsible for SG formation. In this paper we are working within the scenario described in \citet{DErcole2008MNRAS.391..825D,Calura2019MNRAS.489.3269C}. In those models SG forms from AGB ejecta and pristine gas reaccreted from the external medium. SG form in the inner regions as a result of a cooling flow and gas flowing to the inner regions where SG stars are formed. Then, with time the evolution of a cluster erases the spatial differences (e.g. \citet{Tiongco2019} and references within).

This paper is organized as follows. In Section~\ref{sec:NumericalSimulations} there is an introduction to new features which were implemented into \MOCCA code in order to prepare new simulations of GCs with multiple stellar populations (Subsection \ref{sec:MOCCANewFeatures}). Subsection \ref{sec:InitialConditions} provides summary of the \MOCCA models which were computed so far and which are planned to be finished in a near future. In this section we present also new, public version of the \textsc{beans} software, which is a general, web-based solution for distributed data analysis of huge datasets (Subsection~\ref{sec:DataAnalysis}). In the next Section~\ref{s:Results} there is a description of the results obtained from the simulations, together with some general validation tests showing that the current version of \MOCCA code is consistent with the previous version. In the Section~\ref{sec:Discussion} we discuss how the results from the \MOCCA simulations explain (or are consistent) with the already available observations of Milky Way GCs. In particular, we analyze how the findings of this work could help to constrain the initial conditions of the star clusters. In the Section~\ref{s:Conclusions} we summarize our findings about multiple stellar populations, and we present our future plans and projects. In Appendix~\ref{s:AppendixA} we describe how we treat in \MOCCA code the mixed populations and we present briefly the output files. And finally, Appendix~\ref{s:AppendixB} contains an example script which shows how one can analyze in parallel all \MOCCA simulations at once.

\section{Numerical simulations}
\label{sec:NumericalSimulations}

This section presents the new features implemented into \MOCCA code, the initial conditions for the simulations and tools used to perform distributed data analysis.

\subsection{MOCCA -- new features}
\label{sec:MOCCANewFeatures}

The numerical simulations were performed with the \MOCCA\footnote{\url{https://moccacode.net}} code \citep{Giersz1998MNRAS.298.1239G,Hypki2013MNRAS.429.1221H,Giersz2014arXiv1411.7603G}. \MOCCA is currently one of the most advanced codes to perform full stellar and dynamical evolution of real size star clusters. \MOCCA can be used to study various aspects of star cluster evolution, like formation and evolution of compact binaries, e.g. BHs binaries \citep{Hong2020MNRAS.498.4287H}, exotic objects like IMBHs \citep{Giersz2015MNRAS.454.3150G}, blue straggler stars \citep{Hypki2017MNRAS.466..320H}, or can be used to model even the most massive star clusters, e.g. 47~Tuc, \citep{Giersz2011MNRAS.410.2698G}. \MOCCA follows the stellar and dynamical evolution of all stars in the system closely to \textit{N}-body codes and provides almost as much information about single and binary stars \citep{Wang2016MNRAS.458.1450W}. In \MOCCA stellar evolution for single and binary stars was performed with the \textsc{sse/bse} code \citep{Hurley2000MNRAS.315..543H, Hurley2002MNRAS.329..897H} which was strongly updated by \citetBab{Belloni2017a}. Among the most important upgrades are the ones which concern common-envelope phase. Summary of the stellar evolution features one can find in \citet{Kamlah2021}. Strong dynamical interactions in \MOCCA are performed with \textsc{fewbody} code \citep{Fregeau2004-01-004,Fregeau2007ApJ...658.1047F}.

The most noticeable new feature which was added to \MOCCA code, since the first \SURVEYONE \citep{Askar2017MNRAS.464L..36A}, is the support for multiple stellar populations. \MOCCA is now able to follow the full dynamical evolution of multiple stellar populations, and to some extent, also the stellar evolution for different populations. \MOCCA can support up to 10 different stellar populations in one simulation.

A few technical aspects of multiple stellar populations implementation require additional description. Every star in \MOCCA simulation has its own population id (assigned as an initial parameter at time T~=~0). After that, the star can have various interactions with other stars and binaries which might result in a change of its mass (e.g. mass transfers, dynamical mergers). If the star happens to have such an event with a star coming from the same population, the result of the interaction is performed as usual. However, if the interacting star comes from a different population, the star is marked as so-called \textit{mixed population} star. In \MOCCA we currently use stellar evolution codes which do not support mixing between stars with different chemical abundances, thus we are only marking these stars. For events which concern stars from two different populations we take as a current metallicity the one which comes from the more massive star. This simplification allows us to evolve mixed populations stars as normal stars, but one has to keep in mind that the masses, luminosities, colors for mixed populations can be slightly inaccurate. More details on how the mixed population id is build for stars with complex histories is described in Appendix~\ref{s:AppendixA}.

Initial conditions for all \SURVEYTWO simulations were generated using the \textsc{McLuster} code\footnote{\url{https://github.com/agostinolev/mcluster}} \citep{Leveque2021} which is an upgraded version of the original one\footnote{\url{https://github.com/ahwkuepper/mcluster}} \citep{Kupper2011}. Among the most important additions to the original \textsc{McLuster} code is the capability to generate initial conditions for star clusters for up to 10 different stellar populations. Additionally, the initial conditions for the multiple populations are set in such a way that the whole star cluster is in virial equilibrium by solving Jeans equations \citep{Kamlah2021}. Every population can have its own set of initial values. One can choose initial number of single and binary stars for every population, initial model (homogeneous sphere, Plummer, or \citealt{King1966AJ.....71...64K})
. Models can be initially segregated or not \citep{Baumgardt2008}, the stars can be initially fractaled \citep{Goodwin2004A&A...413..929G}, virial ratio can be set to every population separately too. Stellar mass functions can be set to equal masses, \citet{Kroupa2001MNRAS.322..231K} mass function, multi power law (based on \textsc{mufu} by L.~Subr), optimal sampling following the prescriptions of either  \citet{Kroupa2011IAUS..270..141K}, or \citet{Maschberger2012ASPC..453..367M}. Pairing for binary components and semi-major axis distributions can be also set to a few different procedures (e.g. random pairing; ordered pairing; uniform mass ratio (0.1 < q < 1.0) distribution for binaries with component masses larger than $5 M_{\odot}$ according to \citealt{Kiminki2012,Sana2012,Kobulnicky2014} -- selected option for all \MOCCA simulations presented in Section~\ref{sec:InitialConditions}). Eigenevolution can be switched off, can follow \citealt{Kroupa1995aMNRAS.277.1491K}, or can follow a new eigenevolution and mass feeding algorithm (\citealt{Kroupa2013pss5.book..115K}, \citealtb{Belloni2017b}). Concentration parameter can be set up for n-th population with respect to the first population. At last, the metallicity can be also set up separately for every stellar population too. Only the tidal radius can be set to the whole star cluster.

Some details about the output files which are produces by \MOCCA simulations, and which data are being stored, are presented in Appendix~\ref{s:AppendixA}.





\subsection{Initial conditions}
\label{sec:InitialConditions}

We present here \MOCCA simulations performed with the newest version of the \MOCCA code (see Section~\ref{sec:MOCCANewFeatures}). The new simulations form \SURVEYTWO~-- a new major upgrade to the existing \SURVEYONE, which has been successfully used in a number of projects already. \SURVEYTWO consists currently of 257 models.

The initial conditions for the \SURVEYTWO simulations are shown in the Table~\ref{tab:MoccaSurvey2}. Parameter N denotes initial number of bound objects for the first population (FG) and the second population (SG), respectively. A bound object is either a single star or a binary. The parameter $\rm W_0$ is the King model parameter, and can be specified for both populations separately. The parameter $\rm M_{max}$ is the upper mass limit for single stars for both populations separately. For the first population the upper limit is the same -- 150~$\rm M_{\odot}$. For the second population it is 150~$\rm M_{\odot}$ or 50~$\rm M_{\odot}$ (the latter limit mimics a scenario for which the second population contains less massive stars). The next parameter $\rm m_{func}$ is stellar mass functions. For \SURVEYTWO we use two values: the value 0 is for equal mass simulations, and value 1 for \citet{Kroupa2001MNRAS.322..231K} mass function. The next parameter fb is binary fraction for both populations. The number of binaries is defined as $\rm fb_1 * N_1$ for FG and similarly for SG. The parameter $\rm R_{t}$ is the tidal radius for the entire star cluster in parsecs, and $\rm R_h$ is half-mass radius also for the whole star cluster.
The parameter \CONCPOP is the concentration parameter but set to all multiple stellar populations with respect to the FG. It is defined as $\rm R_{hi}/R_{h1}$ -- the ratio between the half-mass radius of the i-th population to the first population. The parameter evol defines whether the stellar evolution is on, off or the stars are point masses. For stellar evolution switched off we assign every star its ZAMS radius to keep stellar collisions and mergers possible. For the case of point-mass stars the radii are not assigned. This parameter is used only for some simulations to compare them with simulations already presented in the literature and for testing.

\begin{table}
\caption[Initial conditions for MOCCA-SURVEY-2]{The table shows selected 
initial conditions of \SURVEYTWO
     -- a set of \MOCCA simulations performed with the upgraded version
     of the code. All of the initial conditions were chosen to test the
     mixing between different stellar populations for star 
     clusters of a real size.
     FG means the first generation, and SG the second one. 
     Some of the parameters are different for these two stellar populations 
     (e.g. N) and some of them are the same in both cases (e.g. 
     $\rm R_h$). The meaning of the initial parameters is
     following:
     $\rm N$ -- initial number of objects; 
     $\rm W_0$ -- King model parameters; 
     $\rm M_{max}$ -- upper mass limit for a single star;
     $\rm m_{func}$ -- stellar mass function;
     $\rm fb$ -- binary fraction;
     $\rm R_{t}$ -- tidal radius [pc]; 
     $\rm R_h$ -- half-mass radius for the whole star cluster; 
     \CONCPOP -- concentration parameter between two populations (e.g. value 0.1 means that SG is 10 times more concentrated than FG); 
     $\rm evol$ -- stellar evolution on/off or point mass (1, 0, -1 
     respectively).
     For details see description in Section~\ref{sec:InitialConditions}.}
\centering
\begin{tabular}{c c c c}
\hline\hline 
Parameter & FG & SG\\ 
\hline
 $\rm N$            & 400k       & 200k, 400k, 600k \\ 
 $\rm W_{0}$      & 6          & 6, 8             \\
 $\rm M_{max}$ [$\rm M_{\odot}$] & 150 & 50, 150       \\
 $\rm m_{func}$   & \multicolumn{2}{|c|}{0, 1}       \\
 \textit{fb} & \multicolumn{2}{|c|}{0.0, 0.1, 0.95} \\
 $\rm R_{t}$ [pc]    & \multicolumn{2}{|c|}{60, 120} \\
 $\rm R_{h}$ [pc]      & \multicolumn{2}{|c|}{0.6, 1.2, 2.0, 4.0, 6.0} \\
 \CONCPOP & \multicolumn{2}{|c|}{0.1, 0.5, 1.0, 1.5} \\
 evol         & \multicolumn{2}{|c|}{-1, 0, 1} \\
 \hline
\end{tabular}
\label{tab:MoccaSurvey2}
\end{table}

The rest of the initial parameters are common for all simulations. 
The semi-major axes distributions follows 
\citet{Kroupa1995aMNRAS.277.1491K},
eigenevolution \citep{Kroupa1995aMNRAS.277.1491K}, reviewed by \citetb{Belloni2017b}, is switched on for \SURVEYTWO by default (with only a few exceptions for some point mass models). The minimum semi-major axis is set up automatically in such a way that it will cause no immediate merger. The maximum semi-major axis is set to 100 AU. Tidal field is due to a point-mass galaxy with mass inside circular cluster orbit and constant rotation velocity equal to 220 $\rm km~s^{-1}$. Metallicity for both populations is set to 0.001 (solar metallicity is Z = 0.02). Mass of neutron stars (NSs) and BHs is determined according to rapid supernova model which includes mass fallback \citet{Fryer2012}. Pair-instability supernovae (PSNs) to entirely disrupt massive stars, and pair-instability pulsation supernovae (PPSNs) are set according to \citet{Belczynski2016A&A...594A..97B}. Electron capture supernovae are also switched on be default for all models. Kick mechanism for NS/BH formation is set to standard momentum conservation (NS and BH have the same kick distributions) and the velocities are set to Maxwellian distribution ($\sigma = 265.0$ km/s -- some BH kick velocities are reduced because of mass fallback \citealt{Fryer2012}). White dwarfs are set not to get any kick velocities upon formation. 

\begin{figure}
\centering
    \includegraphics[width=1.0\linewidth]{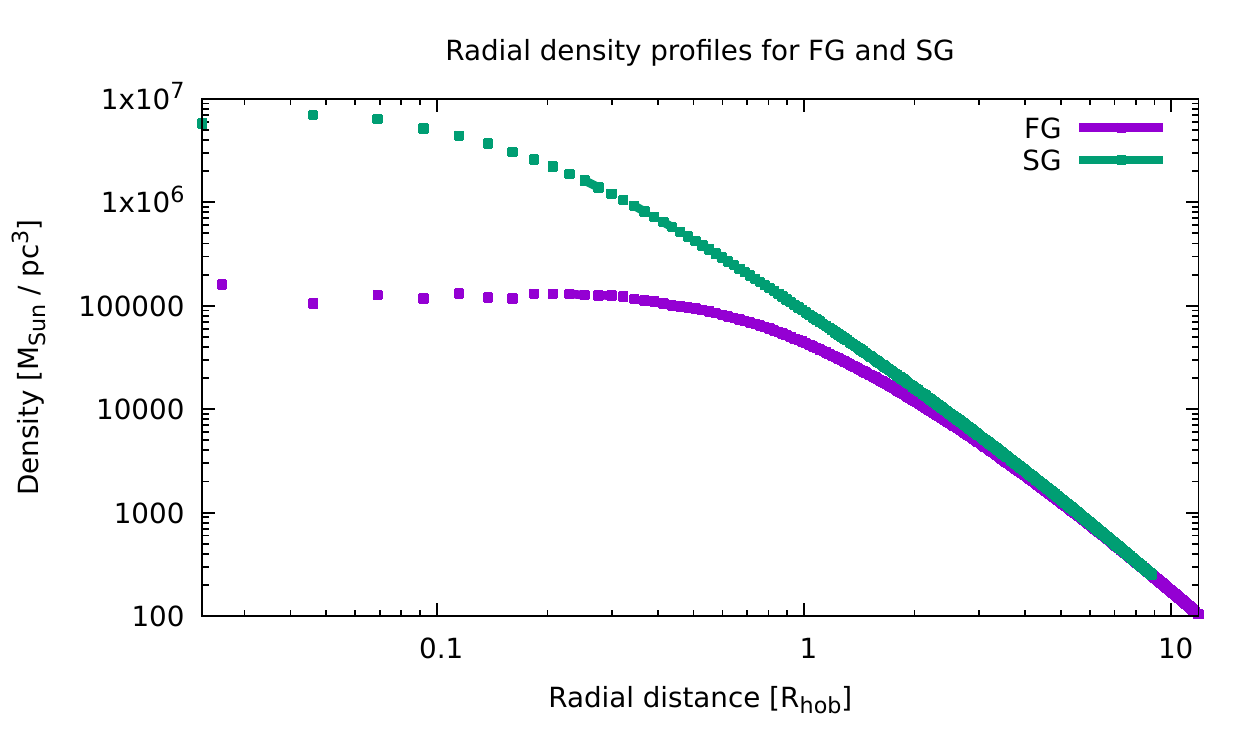}
     \caption{An example radial density profile of FG and SG for one of 
     the \MOCCA simulation ($\rm N_1 = 400k$,
     $\rm N_2 = 400k$, fb = 0.95 
     for both populations, $\rm R_t = 60$ pc, $\rm R_h = 1.2$ pc) as a function of observational half-mass radius (\RHOB). 
     The concentration parameter (\CONCPOP) of the
     SG (green) with respect to the FG (violet) is
     equal to 0.5, which means that the second population is two times 
     more densely concentrated than the FG. }
     \label{fig:SpatialDistributionExample}
\end{figure}

Figure~\ref{fig:SpatialDistributionExample} presents an example spatial distribution of two populations for one of the \MOCCA simulations from \SURVEYTWO as a function of observational half-mass radius (\RHOB). This \MOCCA simulation consists of $\rm N_1 = 400k$, $\rm N_2 = 400k$, fb = 0.95, $\rm R_t = 60$ pc, $\rm R_h = 1.2$ pc, and \CONCPOP = 0.5 (green SG is two times more densely concentrated than violet FG). We include this plot to emphasize what are the scales of the distributions of the two populations. It means that most of the total mass is very much in the center, and the cluster has a lot of space to expand. In many aspects it behaves like an isolated cluster. The plot shows also that SG dominates in the center, whereas FG in the outskirts of the cluster. 

It is important to stress out that the initial conditions were chosen to test the possible mixing between different stellar populations for GCs of realistic sizes. Thus, we have chosen rather dense models in order to make the mixing process more apparent. Moreover, the initial conditions were chosen in such a way that for some of the models FG is denser, for other SG, and on top of that there are some simulations where the distributions of FG and SG are the same. Additionally, we have included also the models in which the SG is more numerous (the sole reason for that was testing, because otherwise it would rather hard to justify why SG could have initially more stars than FG). Moreover, it is important to point out that the paper is mainly focused on the models based on the scenario described in \citet{Calura2019MNRAS.489.3269C}. In this scenario SG forms in the very cluster center, after some time delay (usually a few dozens of Myr), from gas lost due to stellar winds of AGB stars and pristine gas reaccreated by GC during its movement through gas cloud left after GC formation. There are however some deviations in our models to this scenario. We wanted to simplify the physical picture of SG formation. We assumed that there is no time delay for SG formation. Thus, the cluster potential was deeper than for only FG stars. To account for the fast FG cluster expansion (before SG formation) due to strong mass loss of massive stars we assumed larger maximum mass for SG stars. This, a bit artificial assumption, increases overall cluster expansion, keeping it possibly close to expansion of only FG star cluster. Also, because of stronger mass segregation of SG stars, it keeps the SG overall more concentrated respect to the FG as it is suggested by \citet{Calura2019MNRAS.489.3269C}. The hydrodynamic simulations tell us that the SG is relatively dense and concentrated in the very cluster center, so newly formed massive SG stars are prone to collisions and mergers which will produce stars even more massive (in runaway collisions it is probable to form stars with masses larger than even 100 $M_{\odot}$). Our assumption of large maximum IMF mass is a very simple proxy for such massive stars. Additionally, in dense environment, a significant fraction of massive stars can be kicked out of the system due to dynamical interactions before SNe explosions. The large SG fraction, of the order of 1/3, was assumed (together with large maximum IMF mass) to mimic FG (without SG) expansion. The stronger mass loss from SG and stronger energy generation in dense and massive SG help in stronger expanding the cluster.

\begin{figure}
\centering
    \includegraphics[width=0.94\linewidth,right]{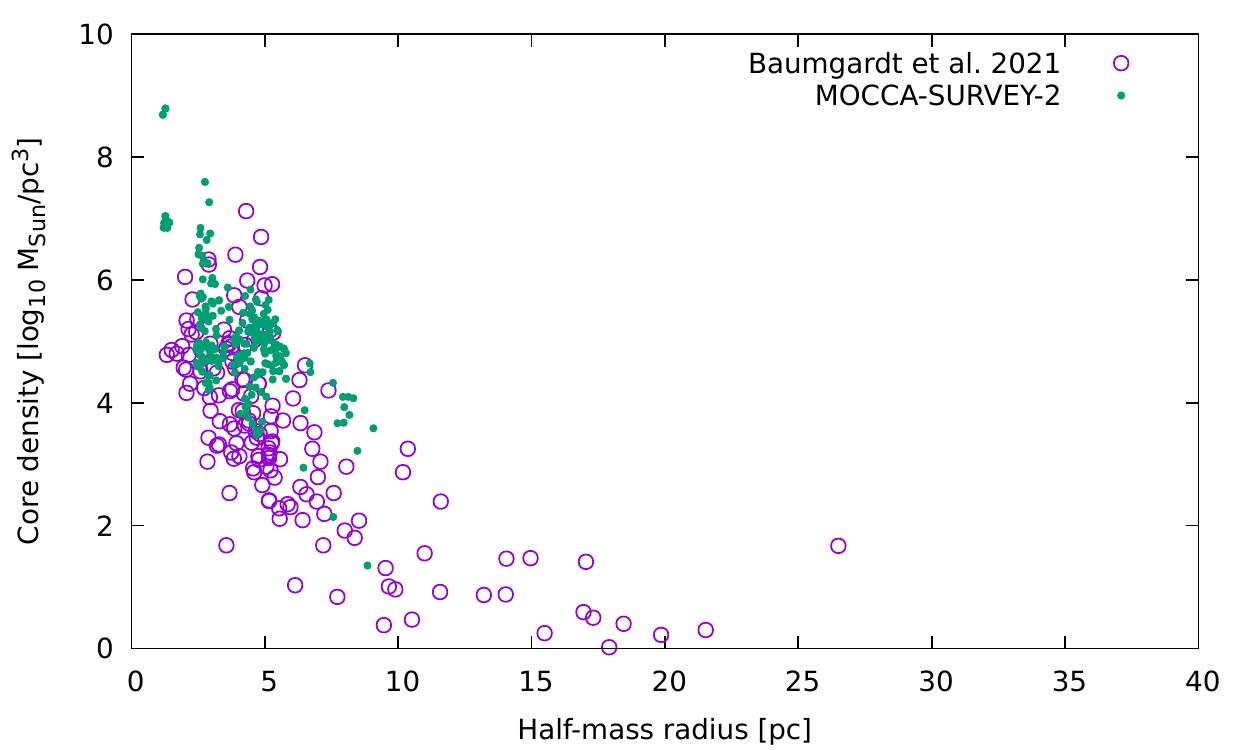}
     \caption{Central density as a function of half-mass radius from \SURVEYTWO simulations for 12 Gyr (green dots) together with the Milky Way GCs parameters taken from \citet[violet circles]{Baumgardt2021MNRAS.505.5957B}. A plot with cluster mass as a function of half-mass radius one can find in \citet{Maliszewski2022}.}
     \label{fig:Holger}
\end{figure}

Although, the purpose of choosing such initial conditions was to test the mixing between populations and we did not plan to simulate real populations of Milky Way GCs it turned out that \SURVEYTWO covers the latter surprisingly well. In Figure~\ref{fig:Holger} we present the global parameters (central density as a function of time, total cluster mass as a function of time one can find in \citealt{Maliszewski2022}) on top panel, and cluster total mass on bottom panel) for \SURVEYTWO simulations at the time 12~Gyr together with the Milky Way GCs parameters determined by \citet{Baumgardt2021MNRAS.505.5957B,Vasiliev2021}. It is interesting to see that even with our initially dense models we cover the real GCs parameters very well. The other parts of the plots, especially these with the higher half-mass radii ($\rm R_h$), are not covered simply because we did not have wide enough initial values (e.g. \SURVEYTWO does not have models with low masses). The initial conditions degenerate over time and the models which initially had different degree of tidal filling can end up at 12 Gyr having similar masses and $\rm R_h$. It is unexpected to see that even starting with quite high densities (of the order of $\rm 10^7 M_{\odot}pc^{-3}$, up to $\rm \sim 10^8 M_{\odot}pc^{-3}$) we eventually get the clusters which cover MW GCs properties quite well (in the given mass regime of our chosen initial conditions). Some of our models have very high initial central densities which may present a challenge to justify them as physically reasonable. However, there are cases where such initial extreme densities are considered as the one which were present at the formation of some globular clusters (e.g. $\rm \approx 10^8 M_{\odot}pc^{-3}$ for $\rm \omega~Cen$, \citealt{Marks2021arXiv211200753M}). The central densities (top panel in Figure~\ref{fig:Holger}) fall especially well within the real GCs values -- it is very unexpected result but gives additional confidence that the findings presented in this work can be applied and discussed with respect to the MW GCs. It has additionally some interesting implications for observations discussed later in the Section~\ref{sec:Discussion}.

The \SURVEYTWO was designed to test a new major version of \MOCCA code. The plan is to extend this survey further with new models. We especially plan to include models which will cover better and more comprehensively the global parameter space of real MW GCs. We also plan to include simulations with gravitational kicks between dynamically merging stars, simulations with \textsc{StarTrack} evolutionary code \citep{Belczynski_2002,Belczynski_2008}, simulations with the use of \textsc{tsunami} code \citep{Trani_2019}, as a replacement of the internal integration code in \textsc{fewbody} \citep{Fregeau2004-01-004}, and simulations with hierarchical systems too. \textsc{tsunami} code is especially important because it includes post-Newtonian terms up to order 2.5 and tidal forces (equilibrium and dynamical tides).

\subsection{Data analysis}
\label{sec:DataAnalysis}


Data analysis for the purpose of this paper is done with \textsc{beans} software\footnote{\url{https://beanscode.net/}} \citep{Hypki2018MNRAS.477.3076H}. \textsc{beans} is open-source, web-based software for interactive and distributed data analysis of huge data sets.

The \MOCCA simulations take a lot of disk space. There are many \MOCCA simulations already done and every one of them contains millions of rows with stellar evolution and dynamical events. Thus, there is a strong need to have a tool for distributed data analysis which would simplify data examination as much as possible. \textsc{beans} helps to solve all these problems. It allows to create self-explanatory, shareable notebooks which provide the full data analysis for the need of this project. 

Access to all \MOCCA simulations are available for all interested parties upon requests on-line through the web page \url{http://beans.camk.edu.pl}. It is our working \textsc{beans} instance which have access to all \MOCCA simulations computed for (older) \SURVEYONE and the current \SURVEYTWO. The web page provides necessary computing power and storage to analyze the \MOCCA simulations in any possible way.

\textsc{beans} is ready for use in production. It already has ability to write scripts in Apache Pig\footnote{\url{https://pig.apache.org}} (high level language for Apache Hadoop platform\footnote{\url{https://hadoop.apache.org}}), AWK, and Python. In order to make it useful for as many users as possible there is a plan to write a number of additional plugins. One of them will be a plugin to work with
Virtual Observatory\footnote{\url{https://ivoa.net}}. In this way \textsc{beans} will gain access to a huge amount of observational data, and to other simulations.

One of the example scripts written in \textsc{beans} is described in details in the Appendix~\ref{s:AppendixB}. It shows how one can analyze all of the \SURVEYTWO simulations in parallel with an easy to understand script written in Apache Pig.

\section{Results}
\label{s:Results} 

The new version of the \MOCCA code was profoundly tested. We have checked a huge number of global parameters of star clusters, snapshots of the system, we compared the results with previous version of the code. We have also performed full technical tests to make sure that there are no problems with the \MOCCA simulations. 

\subsection{Comparison with the \textit{N}-body models}
\label{s:ValidOldVer}

In order to test and to validate the new version of the \MOCCA code it was tested and compared with existing \NBODY models. We have decided to use a few models presented in \citet{Hong2015MNRAS.449..629H,Hong2016MNRAS.457.4507H}, and we are presenting the results for one of them: initial King model $\rm W_0 = 7$, N = 20k, binary fraction = 0.1, $\rm R_{h,FG} / R_{h,SG} = 5$, and the initial binary hardness parameter $\rm \chi_{g, 0} = 20$ (for definitions see Section~2 in \citealt{Hong2015MNRAS.449..629H}). All the particles in the simulation have equal-mass, all the binaries from FG and SG have the same binding energy, and the stellar evolution is switched off. The \MOCCA simulation was started with exactly the same initial conditions (e.g. positions, velocities, initial mass) and with the most up-to-date version to make the comparison most accurate. 

\begin{figure}
\begin{center}
  \includegraphics[width=1.0\linewidth]{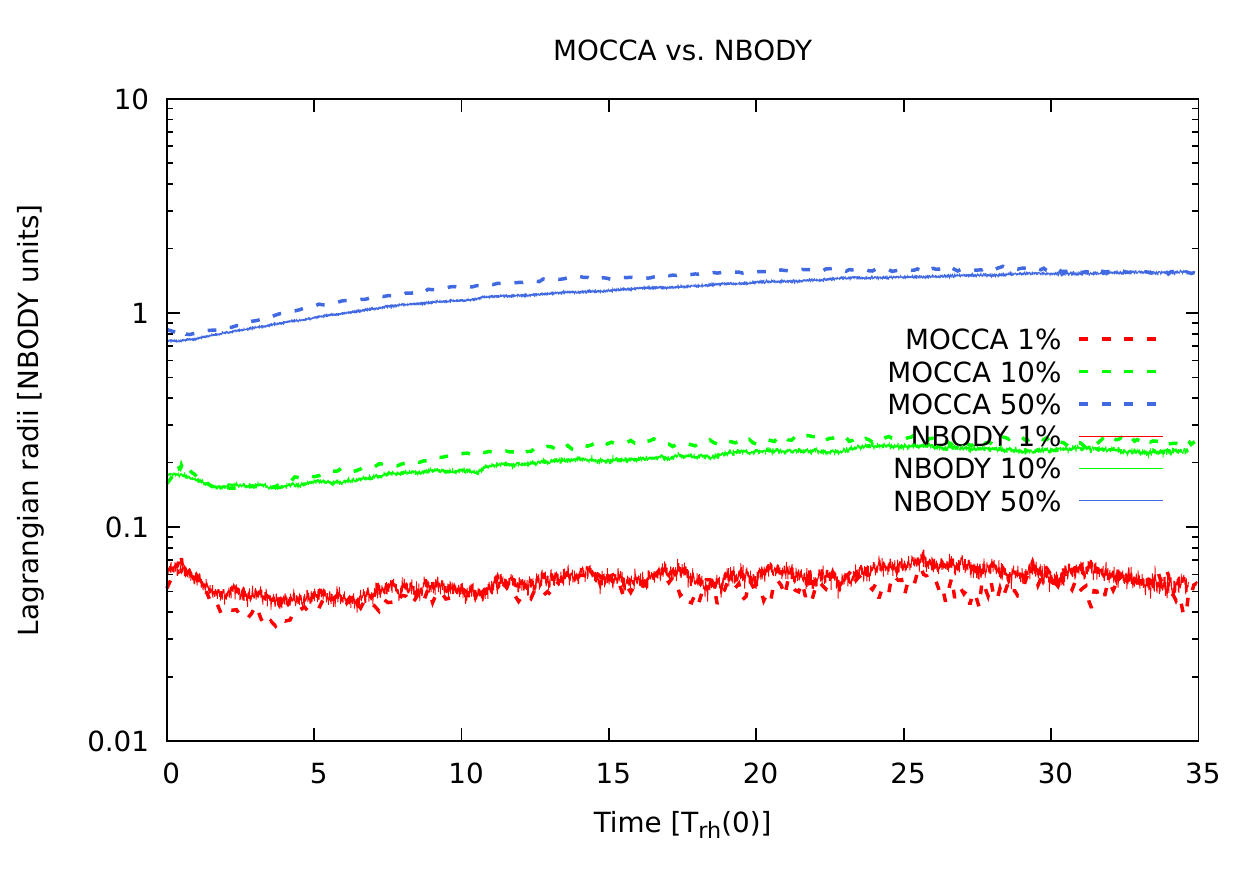}
  \includegraphics[width=1.0\linewidth]{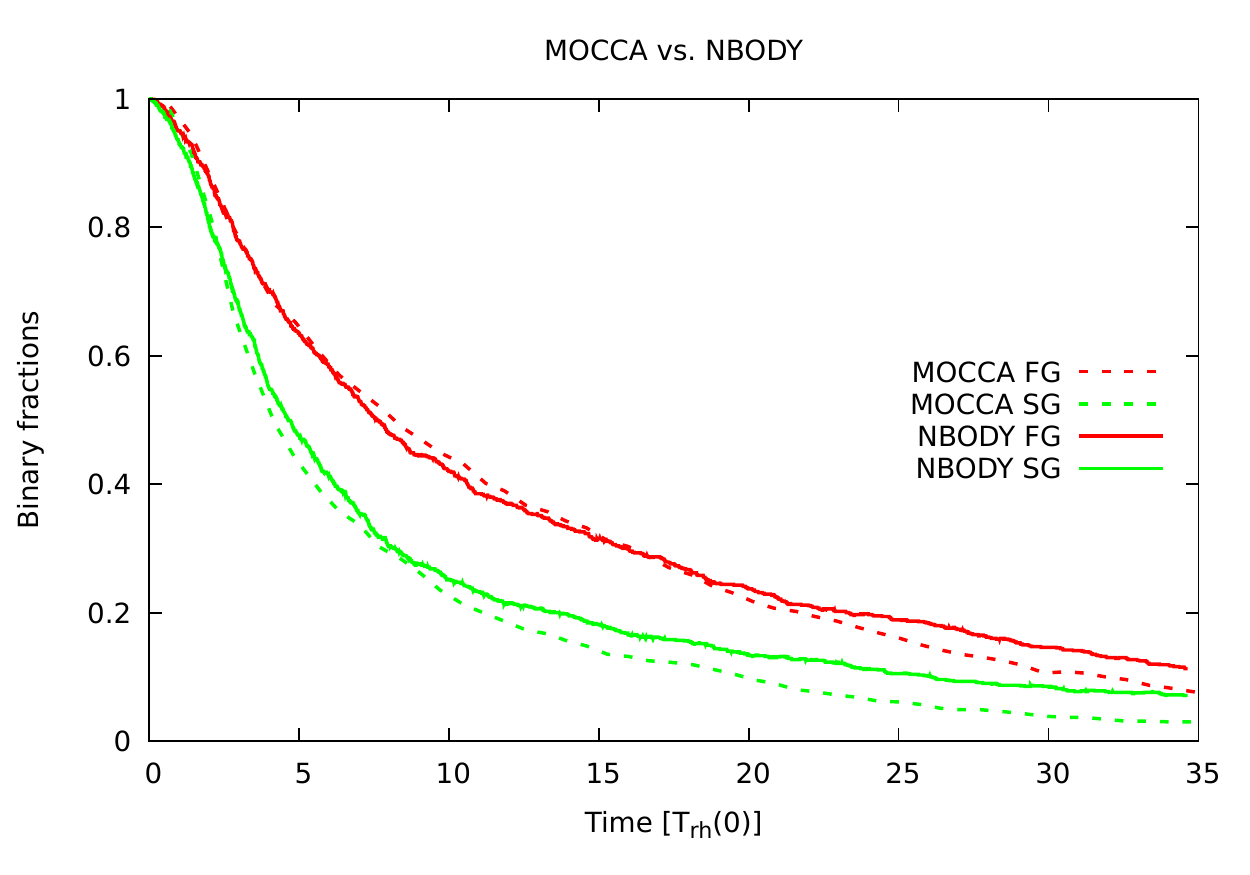}
  \caption{Comparison between one selected simulation from \citealt{Hong2015MNRAS.449..629H,Hong2016MNRAS.457.4507H} (King model $\rm W_0 = 7$, N = 20k, binary fraction = 0.1, $\rm R_{h,FG} / R_{h,SG} = 5$, and the initial hardness parameter $\rm \chi_{g, 0} = 20$, see definitions in Section~2 therein) and \MOCCA for exactly the same initial conditions. Top panel presents Lagrangian radii 1\%, 10\%, and 50\% for both codes for all objects, whereas the bottom panel presents binary fractions for FG and SG for the two codes. Both panels presents the results as a function of time scaled by the initial half-mass relaxation time ($\rm T_{rh}(0)$). For details see Section~\ref{s:ValidOldVer}. }
  \label{f:HongComparison}
\end{center}
\end{figure}

The comparison between results of the one selected \NBODY simulation from \citet{Hong2015MNRAS.449..629H} and \MOCCA run is presented in Figure~\ref{f:HongComparison}. The top panel presents Lagrangian radii 1\%, 10\%, and 50\% in \NBODY units as a function of time scaled by the initial half-mass relaxation times ($\rm T_{rh}(0)$). The comparison is truncated to the maximum time the \NBODY simulation was computed. The bottom panel presents binary fractions for FG and SG for the two codes, also as a function of time scaled by $\rm T_{rh}(0)$. The binary fractions were computed taking into account all bound objects (binaries and single stars) and were scaled by the initial number of objects. The comparison demonstrates that \MOCCA reproduces the global evolution of low-N clusters (N$\sim$20,000) for the different Lagrangian radii (top panel). \MOCCA follows also the evolution of the binary fractions for both populations very accurately (bottom panel). The only difference seems to be present for 50\% Lagrangian radius. This is due to the fact that \MOCCA cannot follow the stage with initial violent relaxation phase as accurately as NBODY. Moreover, the initial model was not in virial equilibrium (see \citealt{Hong2016MNRAS.457.4507H}). \NBODY simulation regained the equilibrium quickly ($\sim$ dynamical scale), which \MOCCA could not follow. The detailed comparison of the older versions of \MOCCA (the Monte Carlo approach) with \NBODY (direct N-body code) one can find in \citet{Giersz2008MNRAS.388..429G} which was done for M67 star cluster, in \citet{Giersz2013MNRAS.431.2184G} which presents comparison with N-body systems up to N = 200k particles, or in \citet{Wang2016MNRAS.458.1450W} where \MOCCA simulations were compared with N-body simulations of four massive GCs with $10^6$ stars and 5 per cent primordial binaries.

\subsection{Multiple stellar populations in tidally filling and underfilling clusters}
\label{s:Imprints} 

Initial conditions for multiple stellar populations are expected to have an influence into the dynamical evolution, in particular mixing, between different populations. Thus, this was taken as the first task to consider while evaluating the results of the \SURVEYTWO simulations. 

We start the description of the results using number of binaries from both populations. These are pure values taken from the \MOCCA simulations, not changed in any way, and degraded by any procedure mimicking the real observations. Later, we will discuss how the found results are influenced by the procedure to selected binaries. We will show whether and if the findings are changed by selected e.g. only main-sequence binaries, observational binaries, or binaries together with single stars. Moreover, the number of binaries, or binary ratios of the two populations are result of complex interplay between the dynamics, binary dissolutions and ejections, which we discuss later in Figure~\ref{fig:NumBinEscDiss}.

\begin{figure*}
\begin{center}
    \includegraphics[width=0.49\linewidth]{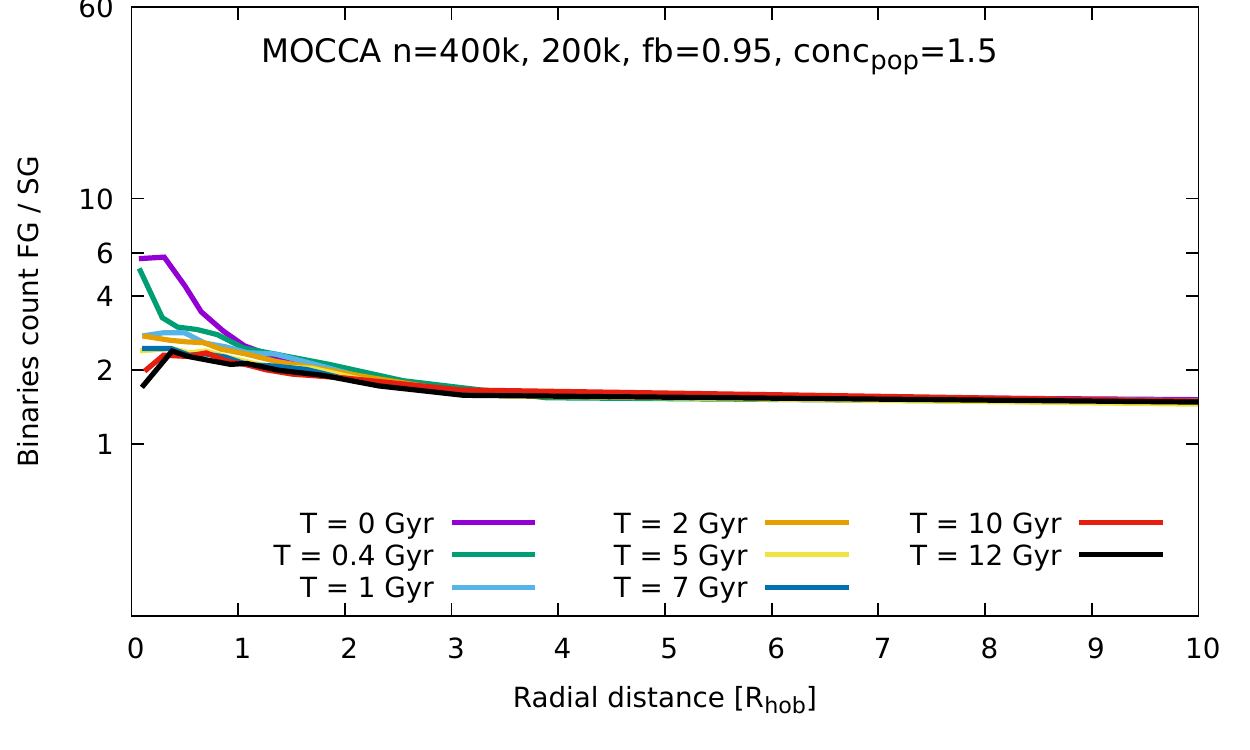}
    \includegraphics[width=0.49\linewidth]{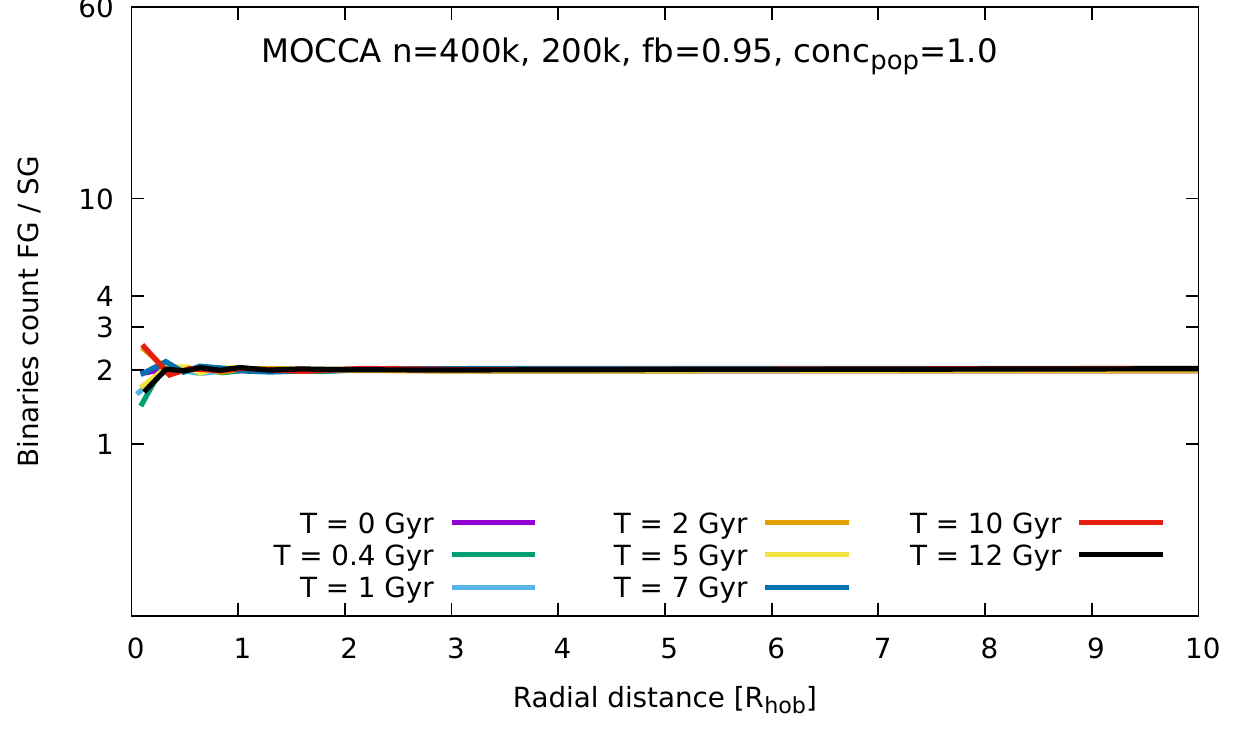}
    \includegraphics[width=0.49\linewidth]{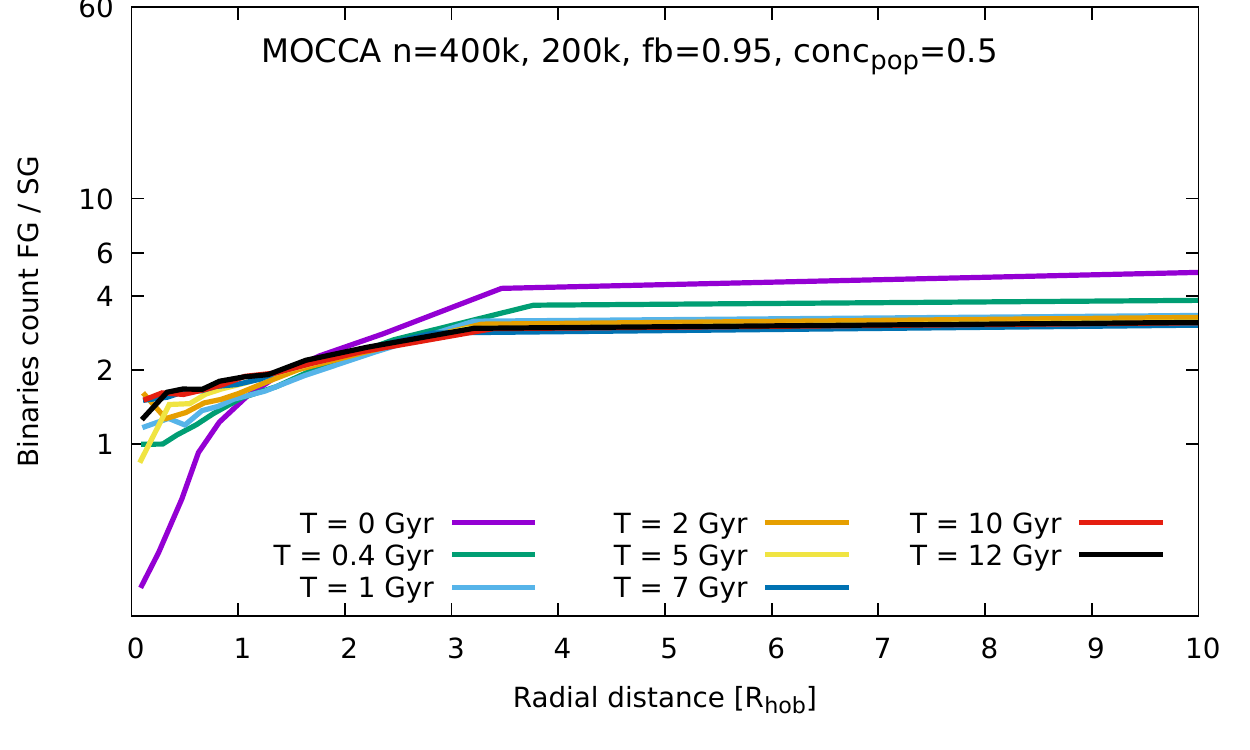}
    \includegraphics[width=0.49\linewidth]{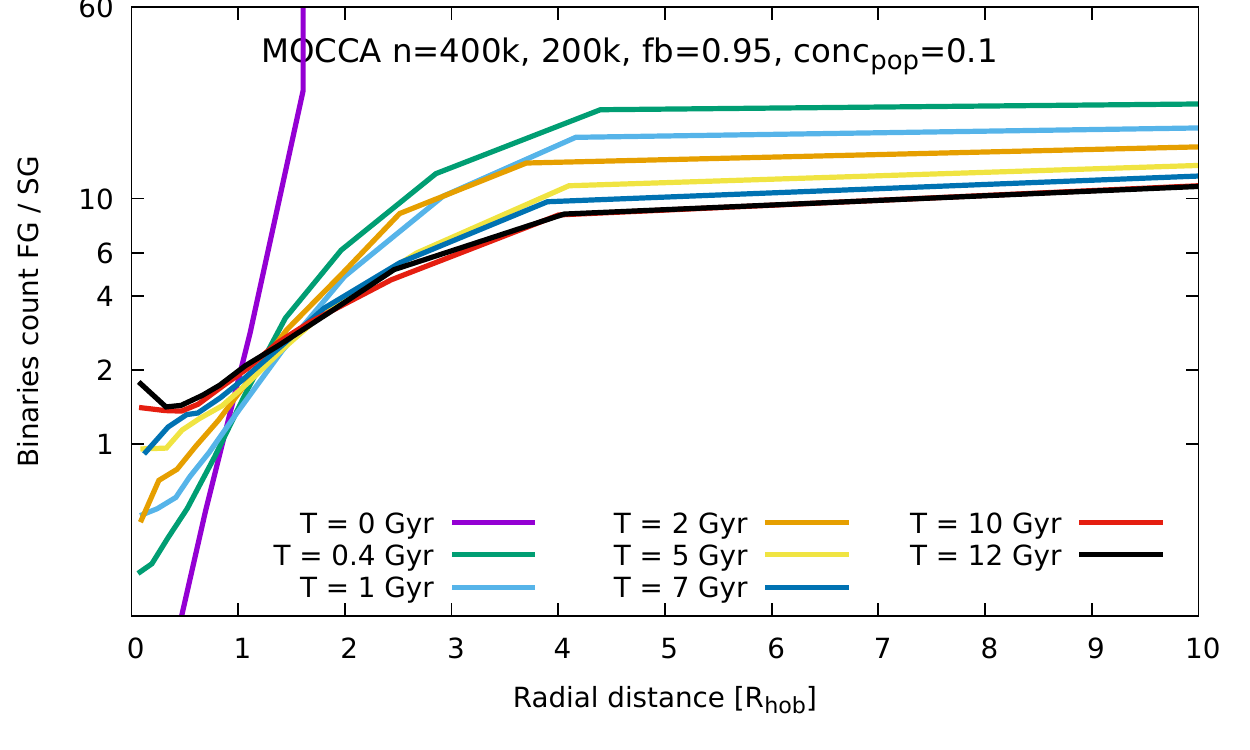}
  \caption{Ratios between all binary stars count from FG to
  SG for four selected \MOCCA simulations, with initial conditions summarized
  in the titles, as a function of radial distance scaled by \RHOB. 
  Each plot consists of eight lines (for different times), and the colors are
  consistent between panels. 
  The only difference between \MOCCA models is the \CONCPOP with
  values from 1.5 (top left), up to 0.1 (bottom right). For details see
  Section~\ref{s:Imprints}.
  }
  \label{f:Mix}
\end{center}
\end{figure*}

Figure~\ref{f:Mix} shows the time evolution of the radial profile of the ratio of the number of FG to SG binaries for a few representative simulations with different initial concentration parameters (\CONCPOP).
All \MOCCA simulations presented in this figure start with the same initial conditions and differ only in the \CONCPOP parameter. In our case, having only two stellar populations, \CONCPOP defines the ratio between the half-mass radius of the SG to the half-mass radius of the FG. The highest \CONCPOP = 1.5 is for the \MOCCA simulation presented on the top left plot, the lowest \CONCPOP = 0.1 is on the bottom right plot. The plots are ordered by decreasing \CONCPOP values. It is important to stress out that the only parameter different for these four simulations is just the concentration parameter between two populations. The main purpose of this figure is to show, in general, how the ratio between the number of binaries from FG and SG evolves with time for different concentration parameters. Unless other specified, every \MOCCA simulation presented since this place, have the initial condtiions: N = 400k, 200k, $\rm W_0 = 6, 6$, $\rm M_{max} = 150, 150 M_{\odot}$, $\rm m_{func} = 1, 1$, fb = 0.95, 0.95, $\rm R_t = 60$, and evol = 1.

The two top panel in Figure~\ref{f:Mix} are included here only for testing because \CONCPOP = 1.5 (spatial distribution of SG is more extended than FG), and \CONCPOP = 1.0 (FG, and SG have the same initial spatial distribution) are hard to justify as a real scenario for the SG formation. However, the top left panel shows that if the FG is more concentrated then it naturally loses more binaries as a result of dynamical interactions, thus the ratio drops with time and eventually at Hubble time drops closer to 1.0. Also, for \CONCPOP = 1.5 the ratio between binaries count for the distances larger than $\rm R_{hob} = 1$ does not change significantly over time (there is no apparent mixing happening in the outskirts of the cluster). The top right panel shows that if the both populations start with the same spatial distribution there is actually no change over the Hubble time for the FG / SG ratio. It is expected behavior which shows that \MOCCA code is working correctly. The slightly scattered values in the center are simply the fluctuations. 

For the \MOCCA simulations with deeper concentrated second populations (two bottom plots) one can see that there appears a very distinct feature. The ratio between the number of binaries from FG to SG for the central regions below \RHOB (i.e. values on the X axis smaller than 1.0) for the beginning of the simulations (e.g. 400~Myr, 1~Gyr) have small values. This is because initially the second population is more deeply concentrated. Then, with time this ratio is getting larger, and eventually it becomes larger than one. This means that in the central Lagrangian radii there are now less binaries from SG. These binaries, initially more concentrated, are being destroyed or ejected. Eventually, the ratio throughout all Lagrangian radii is larger than 1.0 -- FG gets more numerous than SG for the entire cluster in general agreement with the findings of \citet{Hong2015MNRAS.449..629H,Hong2016MNRAS.457.4507H,Vesperini2011MNRAS.416..355V,Sollima2022}.

Another distinct feature observed in Figure~\ref{f:Mix}, for $\rm conc_{pop} < 1.0$ models, is that the ratio is getting larger for the central regions (essentially below the half-mass radius) and is getting lower for the outskirts of star clusters (further than \RHOB). Initial shape of the ratio between FG and SG flattens for the entire cluster which means that there is mixing in all regions of the star cluster. 
For \CONCPOP = 0.1 the initial ratio $\sim 0.3$ (blue color) at 400~Myr increased to $\sim 2.0$ at 12~Gyr for the central regions and dropped from $\sim 15.0$ to $\sim 8.0$ for the outskirts of the star cluster ($> 4 \times$\RHOB).

\begin{figure*}
\begin{center}
    \includegraphics[width=0.49\linewidth]{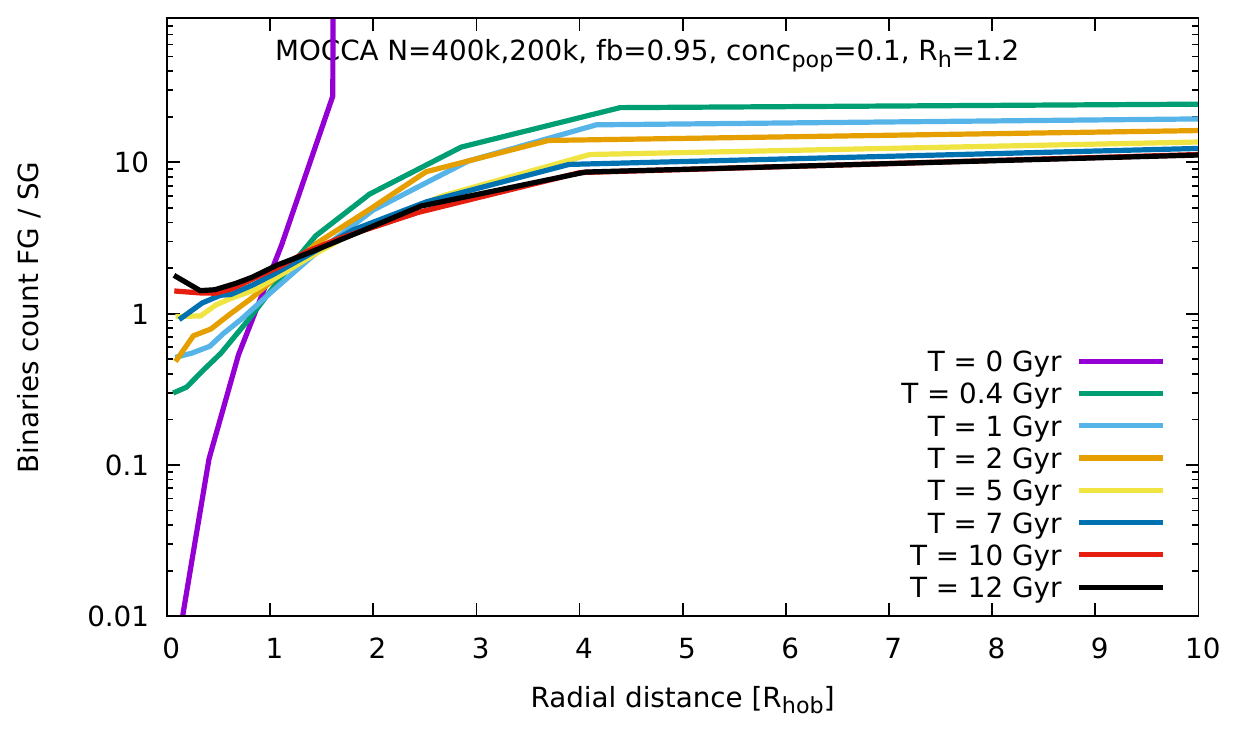}
    \includegraphics[width=0.49\linewidth]{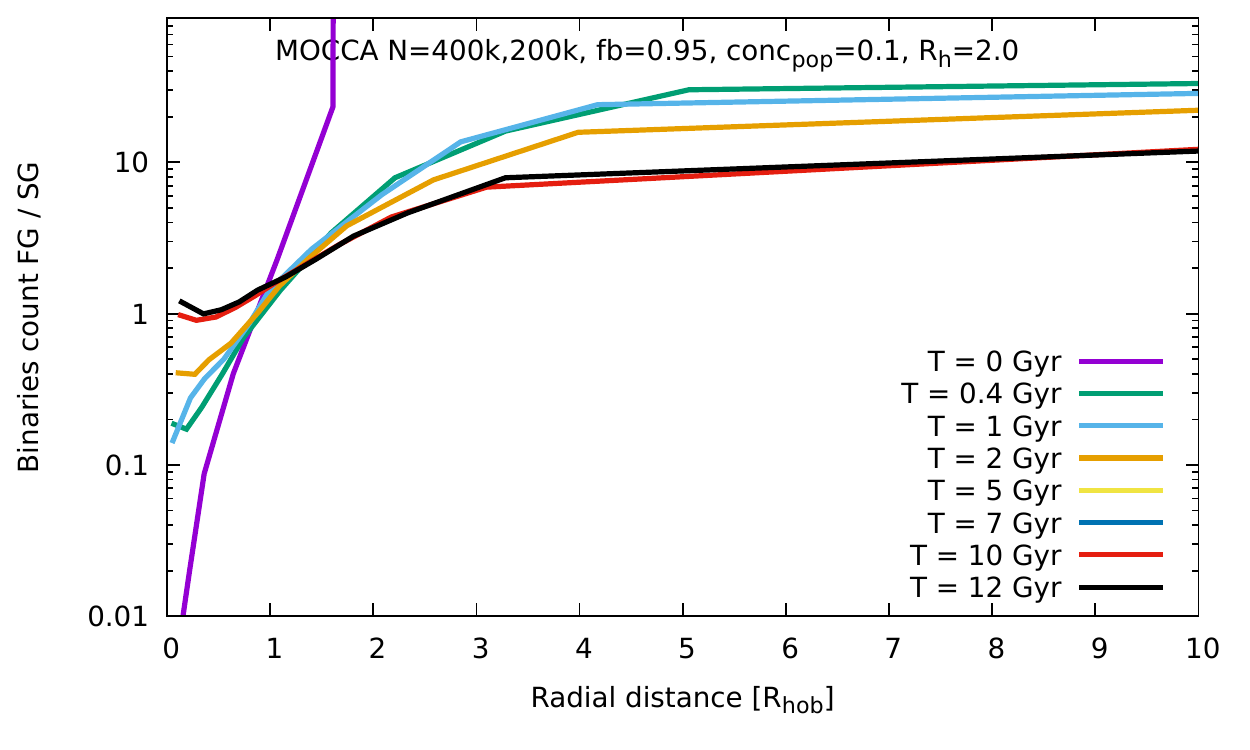}
    \includegraphics[width=0.49\linewidth]{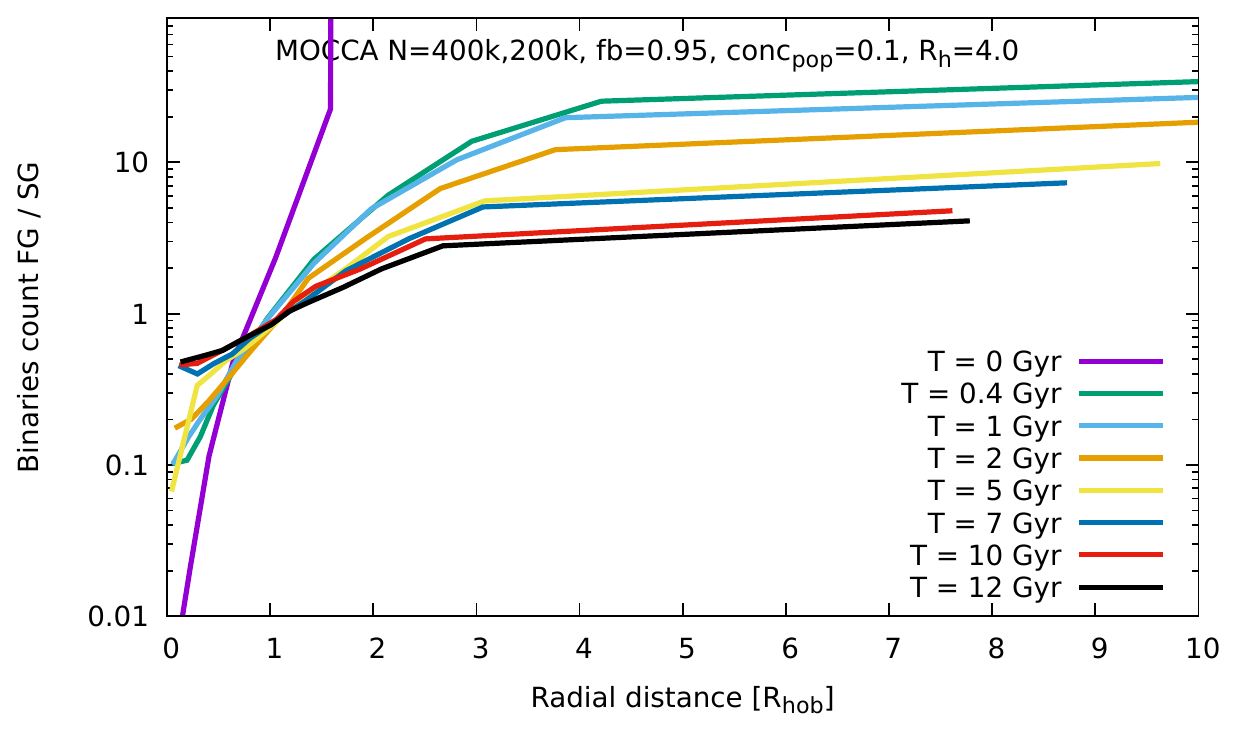}
    \includegraphics[width=0.49\linewidth]{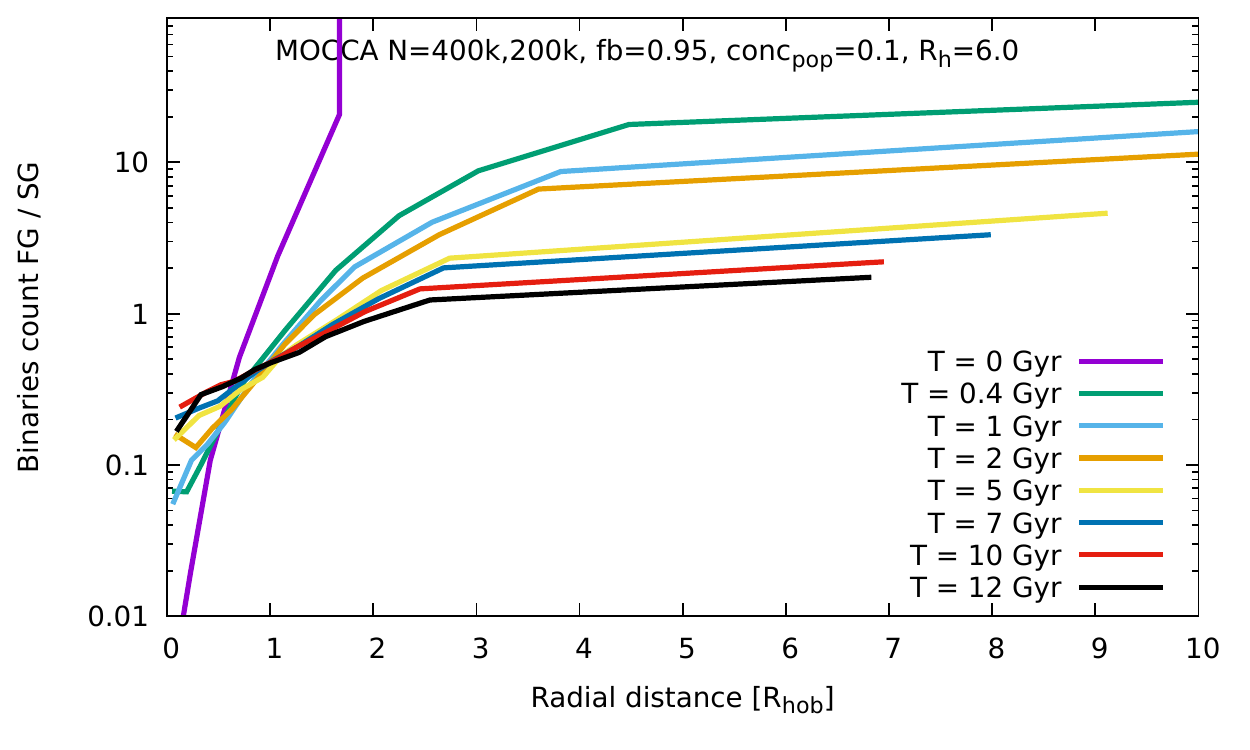}
  \caption{Description as in Figure~\ref{f:Mix}. The \MOCCA models differ
  only with $\rm R_h$. For details see text.
  }
  \label{f:MixDiffRh}
\end{center}
\end{figure*}

The two bottom plots in Figure~\ref{f:Mix} reveal interesting time evolution of the binary ratio for the two populations for two models which differ in by how much the SG is more deeply concentrated with respect to FG. We have decided to investigate it more deeply and in Figure~\ref{f:MixDiffRh} we present also the binary ratios between FG and SG but for four models which have the same initial conditions, and only different $\rm R_h$ (from 1.2 pc which is tidally underfilling model, up to 6.0 pc which is model close to be tidally filling). The overall evolution of the ratios for times from T = 0, up to the Hubble time present similar evolution. At the time T = 0 SG is 10 times more deeply concentrated than FG (\CONCPOP = 0.1). With time, this ratio increases for all models but in principle only in the central regions of the cluster (where dynamical interactions play much more significant role). In the outer regions of the cluster this ratio drops for all models. However, there is a number of interesting differences for tidally filling and underfilling clusters. For tidally underfilling models (e.g. $\rm R_h = 1.2$ pc) the binary ratio FG / SG gets larger than 1.0 after a few Gyr. This means that for such dense models, where SG was deeply concentrated and more numerous in the center, there is a large number of SG binaries which were destroyed or ejected from the center. In such models SG population quickly burns out and as a result at the Hubble time the ratio is larger than 1.0 for the entire star cluster. In turn, for the models which are tidally filling (e.g. $\rm R_h = 6.0$ pc), the initial ratio in the center is also much smaller than 1.0 (SG dominates in the center). In such models SG are also being burned by dynamical interactions, however on much less scale. Thus, with time the ratio increases (there is more and more SG being destroyed), but the ratio does not go above the value 1.0. For tidally filling clusters, the SG still dominates in the center of the GC, whereas in the outskirts of the GC, the FG dominates. Additionally, for tidally filling clusters the binary ratios FG / SG decreases more significantly in the outskirts of the cluster. This is a consequence of the fact that tidally filling clusters have more FG binaries closer to the tidal radius and lose a larger fraction of FG single and binary stars.

The ratio between FG to SG binaries during 12 Gyr of evolution changes much more profoundly for tidally filling clusters (bottom right panel in Figure~\ref{f:MixDiffRh}). The ratio at 0.4 Gyr are of the order of 20.0 at a distance of several \RHOB, and after Hubble time the ratio drops to a value just slightly larger than 1.0 for our model. For tidally filling models SG is burning in the center, whereas FG are preferentially removed or destroyed in the cluster outskirts -- this changes the mass function \citep{Vesperini2021MNRAS.502.4290V}. For tidally underfilling clusters (top left panel in Figure~\ref{f:MixDiffRh}) the ratio changes less significantly. Such relation could be a valuable tool for observations to try to infer some boundaries for the initial conditions of the GCs.

\begin{figure}
\begin{center}
    \includegraphics[width=1.0\linewidth]{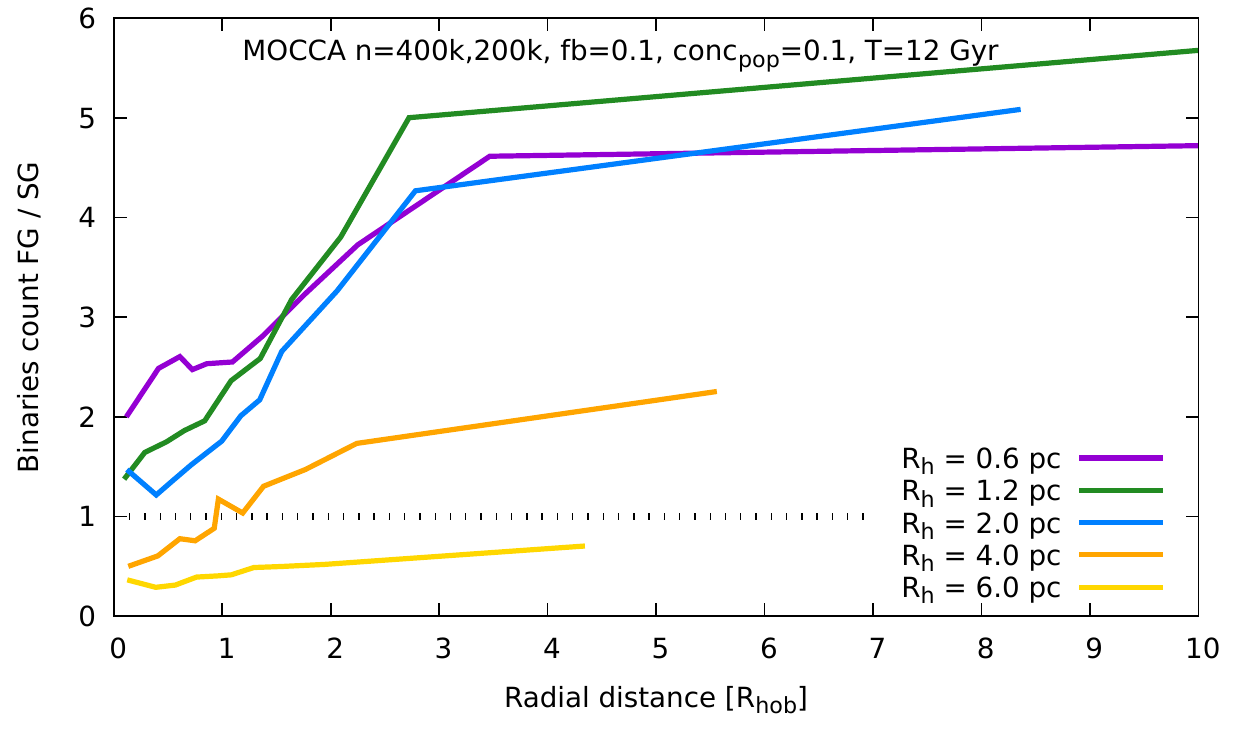}
    \includegraphics[width=1.0\linewidth]{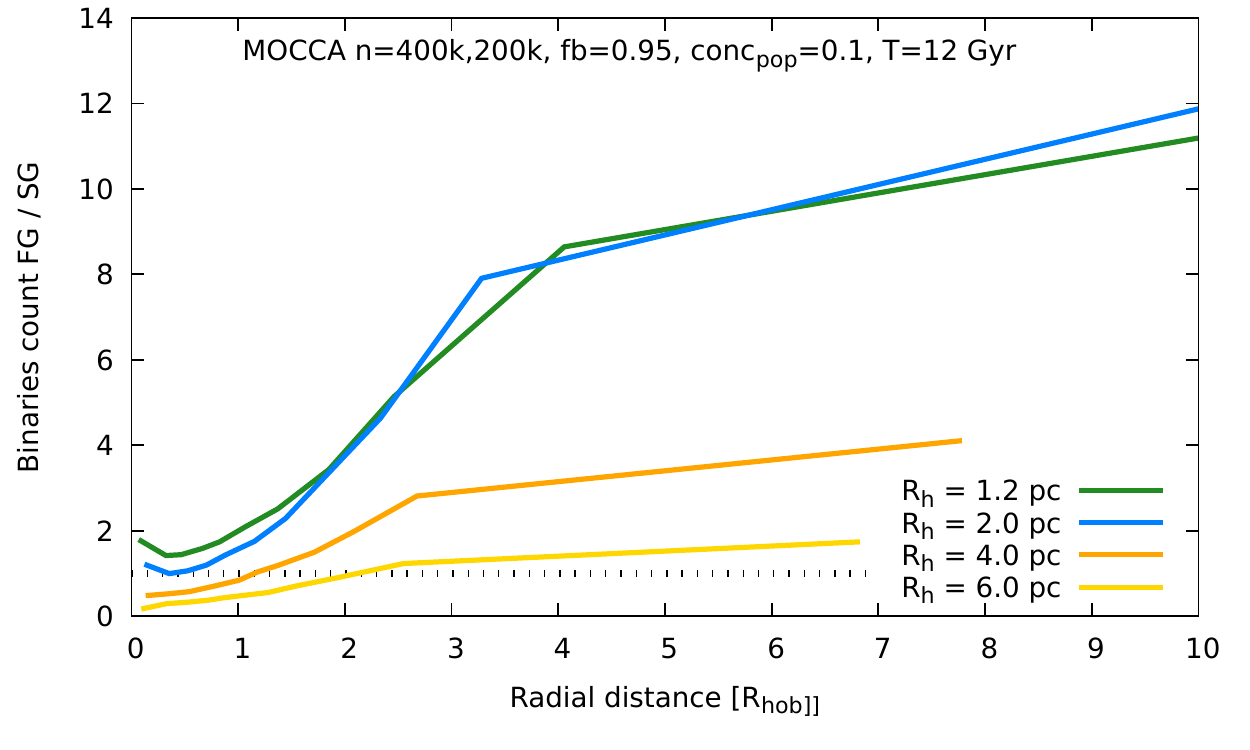}
  \caption{Ratios between all binary stars count from FG to
  SG for a few selected \MOCCA simulations for the time 12~Gyr for different
  initial half-mass radii ($\rm R_{h}$) as a function of \RHOB. 
  Top panel contains \MOCCA
  simulation with initial 10\% of binaries, the bottom panel 95\%. Other
  initial parameters are summarized in the captions. The \CONCPOP between
  both populations is the same for all models and equals 0.1 (SG is 
  initially ten times more concentrated than FG). Additionally, artificial line at value 1.0
  is plotted as a reference, dividing a plot into two regions: above the line
  the number of FG binaries dominates, below FG binaries. See 
  Section~\ref{s:Results} for details.}
  \label{fig:MixingOnly12Gyr}
\end{center}
\end{figure}

One of the most prominent feature of the evolution of ratios between binaries from FG to SG are for times T~=~12~Gyr (black curves in Figure~\ref{f:Mix}, and Figure~\ref{f:MixDiffRh}). They reveal how the ratio behaves for tidally filling and underfilling clusters. In order to investigate this more closely Figure~\ref{fig:MixingOnly12Gyr} shows the ratios between binaries FG / SG as a function of radial distance (scaled by \RHOB). All the ratios come from T~=~12~Gyr and were collected from a set of \MOCCA simulations which differ only in the initial half-mass radii (the smaller it is the more tidally underfilling and more dense the cluster is initially). On the top panel in Figure~\ref{fig:MixingOnly12Gyr} there are \MOCCA simulations for 10\% of initial binary fractions, and the bottom panel for 95\%. The rest of \MOCCA initial conditions are summarized in the captions of the plots (notice that all of the simulations have \CONCPOP equal to 0.1 which means that SG is ten times more concentrated than FG). 
On both panels a dashed black line at value 1.0 is plotted for reference. It represents radial distances at which the number of binaries from FG equals SG. All points above this line denotes that there are more FG binaries, and all points below the line represents radial distances at which the SG is more numerous. All of the \MOCCA simulations in Figure~\ref{fig:MixingOnly12Gyr} have initially more numerous FG than SG by factor of two. The figure shows that the simulations for which FG, at the time T~=~12~Gyr, is still more numerous is only for the clusters which initially were tidally underfilling (smaller $\rm R_h$ values). Deeper investigation of the internal structure of these models revealed that the clusters, because of their high initial densities, evolve very quickly towards core collapse, and after that the cluster just expands nearly homogeneously and continuously with time (all Lagrangian radii increase in the same way with time, even 1\% Lagrangian radius). For these clusters there was plenty of space to expand and preserve the more numerous FG. Only in the central regions the ratio FG to SG is significantly lower, but yet still larger than 1.0. For tidally underfilling models the initial ratio is well below 1.0 (see violet curve on top left panel in Figure~\ref{f:MixDiffRh}). SG are significantly more numerous at the time T = 0, but because of the high density SG are quickly burned out and eventually for tidally underfilling clusters FG gets more numerous even in the central regions of the cluster. With larger $\rm R_h$ values (tidally filling models), the ratio between FG and SG binaries looks different. At some point (in our models $\rm R_h > 2.0$ pc) SG gets more numerous at the time 12~Gyr at least for the central regions of the cluster. Initially, the number of SG binaries is dominant only in the central regions (for $\rm R_h = 4.0$ pc for radial distances less than one $\rm R_h$), whereas in the outskirts of the star cluster FG is still more numerous. But at some value (in our case $\rm R_h = 6.0$ pc), the cluster is initially tidally filling and that changes the resulting number of FG binaries in comparison to SG. For such a model $\rm R_h$ increases because of the initial mass loss of the massive stars, and then because of the energy which is being generated in the core. In turn, $\rm R_t$ decreases due to mass loss and ejections of stars. The cluster loses more and more FG binaries, because they are closer to the tidal field, and eventually the number of SG binaries begins to dominate on all radial distances.


\begin{figure*}
\begin{center}
\includegraphics[width=0.49\linewidth]{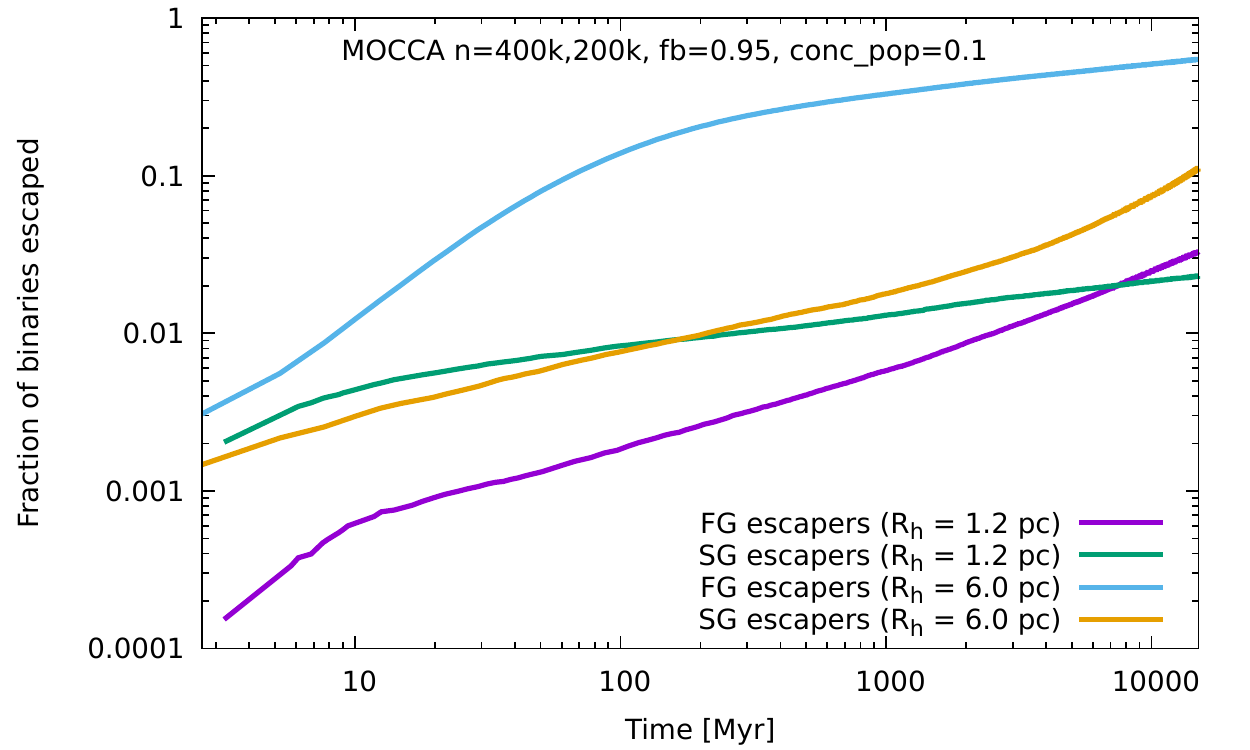}
  \includegraphics[width=0.49\linewidth]{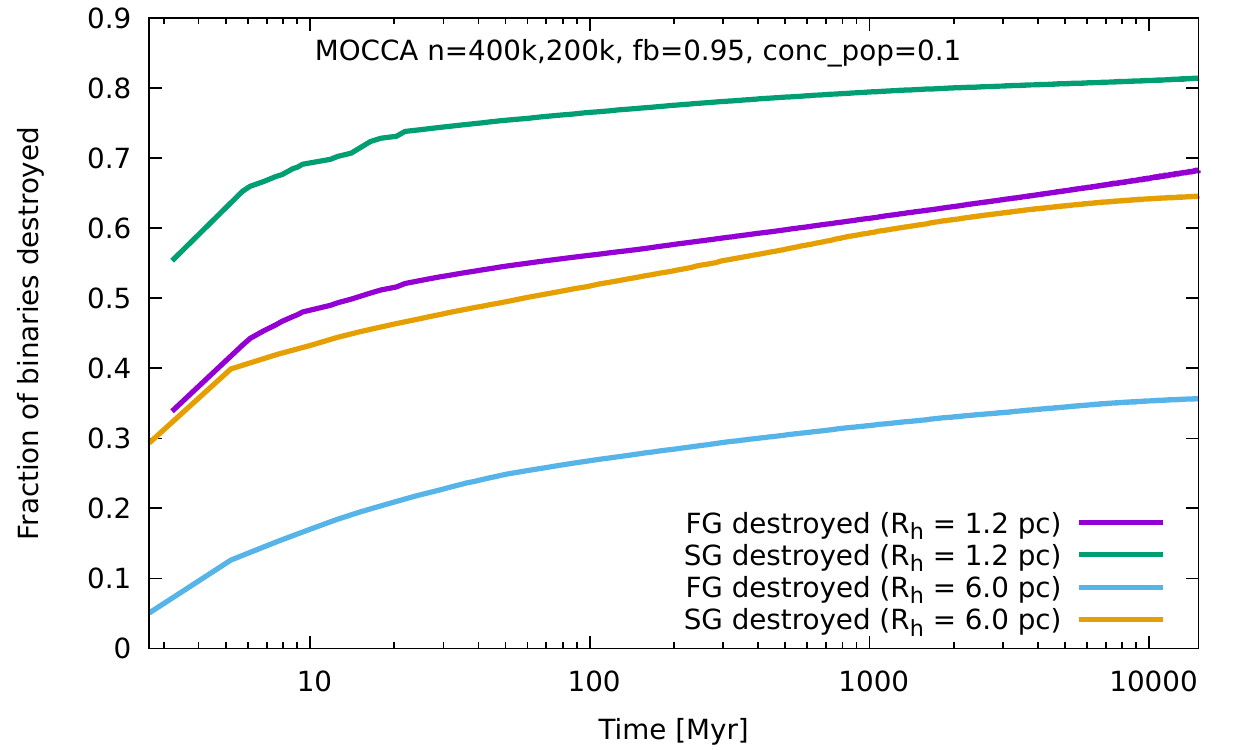}
  \caption{Number of binaries escaped from a star cluster (left panel) 
  and number of binaries dissolved due to dynamical interactions and 
  as a result of stellar evolution (right panel) for FG and SG binaries 
  for two selected \MOCCA simulations as a fraction of the initial number of binaries. \MOCCA simulations 
  have parameters summarized in the titles, but the only 
  difference between them is $\rm R_h$ 
  with values 1.2 -- tidally underfilling cluster, 
  and 6.0 -- tidally filling cluster. For every \MOCCA simulation there 
  are two lines on each panel for FG and SG binaries. For details see
  Section~\ref{s:Imprints}.
  }
  \label{fig:NumBinEscDiss}
\end{center}
\end{figure*}

Figure~\ref{fig:MixingOnly12Gyr} shows that the ratio between the number of FG and SG binaries presents the same features for 10\%, and 95\% initial binary fractions. In the center of clusters the ratio is smaller than 1 ($\rm R_h$ = 4.0, 6.0), and in the outskirts the ratio is large for both of the binary fractions. It seems that the results presented in the Figure are not influenced significantly by the initial binary fraction and the ratios between FG and SG behave similarly.

Figure~\ref{fig:NumBinEscDiss} shows, as an example, the number of escaped (left panel) and destroyed binaries (right panel) for the two populations in two \MOCCA simulations (one underfilling -- $\rm R_h = 1.2$ pc, and one filling -- $\rm R_h = 6.0$ pc) as a fraction of the initial binaries number for any given population. The Figure explains why the number of binaries changes like e.g. in Figure~\ref{fig:MixingOnly12Gyr} differently for filling and underfilling clusters. For the model tidally underfilling the fraction of destroyed binaries is several times larger than for the model tidally filling (SG is being destroyed the most, and at the time 12 Gyr 80\% of all SG binaries are eventually dissolved that way). In turn, for the model tidally filling, the more important mechanism for removal of binaries is due to escapers (around 0.5 of FG binaries escaped after 12 Gyr, and only 0.1 of the SG -- this is because more FG binaries are less concentrated than the SG).  There are two additional features visible in Figure~\ref{fig:NumBinEscDiss}. The ratio between destroyed FG and SG binaries is noticeably larger for the tidally filling model than for the underfilling one. Also, what is interesting and unexpected, the number of escaped binaries is larger for SG than FG for the tidally underfilling model. This suggests that escapes connected with strong dynamical binary interactions play an important role in the binary removal process.

\begin{figure}
\begin{center}
    \includegraphics[width=1.0\linewidth]{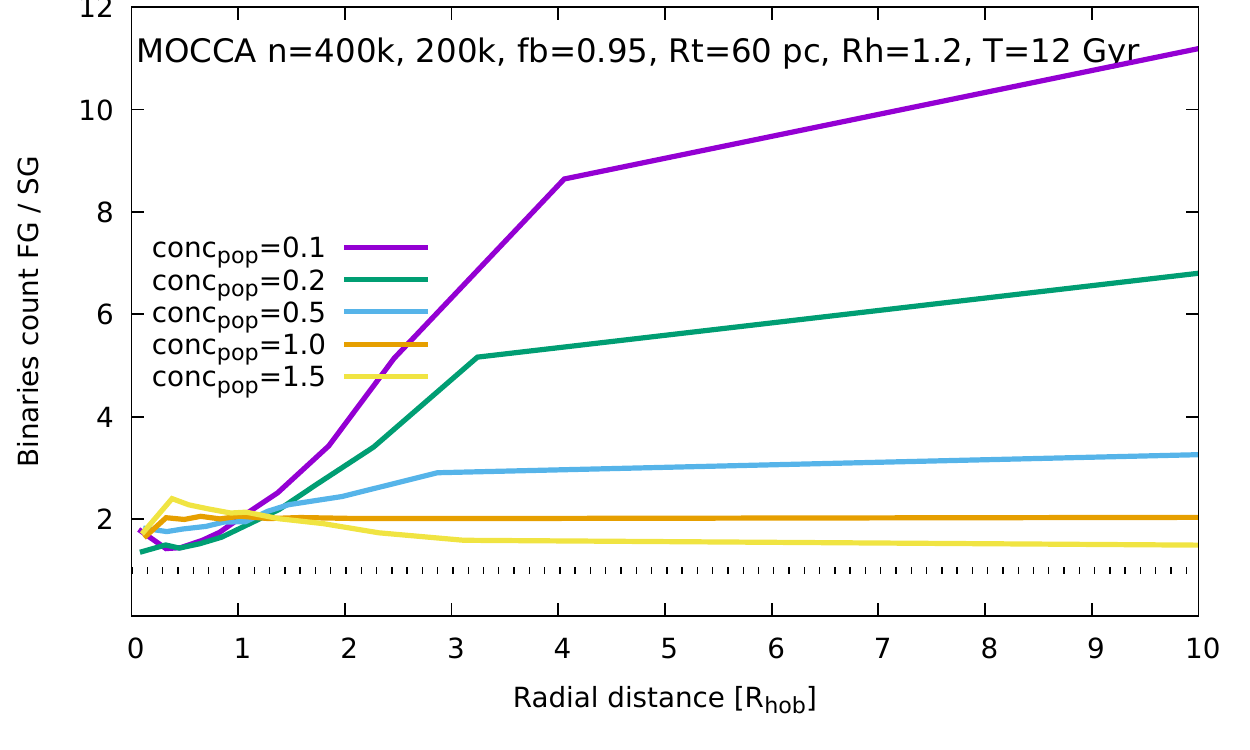}
  \caption{The ratio between binaries count of FG and SG for a set
  of \MOCCA simulations which differ in \CONCPOP values only (e.g. 
  $\rm conc_{pop} = 0.1$ means that SG is initially ten times more 
  concentrated than FG) as a function of radial distance scaled by \RHOB. The other
  initial parameters are summarized in the caption. Additionally, artificial
line at value 1.0 is plotted as a reference (above the line the number of FG binaries dominates). The model with 
  $\rm conc_{pop} = 1.5$, in which SG is initially less concentrated 
  than FG, is rather hard to explain physically. However, it is 
  presented here for completeness (\SURVEYTWO is a test survey and 
  some of the models had such concentrations for SG, see Section~\ref{sec:NumericalSimulations} for details).}
  \label{f:MixRhConst}
\end{center}
\end{figure}

For completeness let's examine the behavior for the ratio between binaries from FG to SG also from the point of view of the \CONCPOP parameter. Figure~\ref{f:MixRhConst} shows the ratio between binary stars FG to SG at the time 12~Gyr, but the simulations in this case differ only by the \CONCPOP parameter. All other initial parameters are the same and summarized in the caption. All of the \MOCCA models in this figure are underfilling ($\rm R_h = 1.2$~pc). The figure clearly shows that for the smaller \CONCPOP values we observe the same result for the ratio between binaries count FG to SG like in the previous Figure~\ref{fig:MixingOnly12Gyr}. Initially the overall number of FG is larger than SG for the whole cluster, but not in the central regions. Because SG was initially more concentrated the ratio was below value 1.0 in the center. With time, and because SG binaries were quickly dissolved or ejected, this ratio reverts and as a result the FG / SG ratio gets larger than 1.0 even for the central regions of the cluster. Figure~\ref{f:MixRhConst} shows that the more the SG is concentrated, the larger the ratio between FG to SG is. This can have an important implications for observations. For \CONCPOP equal to 1.0 (FG is initially distributed statistically in the same way as SG), the ratio is rather flat around value 2.0, so in principle the ratio between initially more numerous FG populations is preserved. There is only a slight bump in this ratio for the central regions (below $\rm R_{hob} < 0.5$ pc) which is connected with fluctuations.

The model, which initially consists of less concentrated SG stars ($\rm R_{h SG}/R_{h FG} = 1.5$) is shown here for completeness despite the fact it is rather hard to justify initial concentration (there is no reason why SG could be less concentrated than FG). However, it is interesting to see that for such a model the initial radial distribution of FG and SG are basically saved. FG is more numerous in the central regions, whereas SG is more numerous in the outskirts of the cluster. After 12~Gyr this distribution is still preserved and binaries count reverts from a value $>2.0$ to a value $<2.0$ indeed around \RHOB. Here again, there is just small decrease in the number of FG in the very center ($\rm R_{hob} < 0.2$) which is connected with FG being preferentially more often dissolved or ejected due to dynamical interactions.

\subsection{Different ways of computing binaries count for FG and SG}
\label{s:DiffWays}

We have checked whether the changes of the ratio FG / SG (e.g. as in Figure~\ref{f:Mix}) are also visible for the binaries if we take into account binaries computed in a different way e.g. only MS binaries. 

\begin{figure}
\begin{center}
  \includegraphics[width=1.0\linewidth]{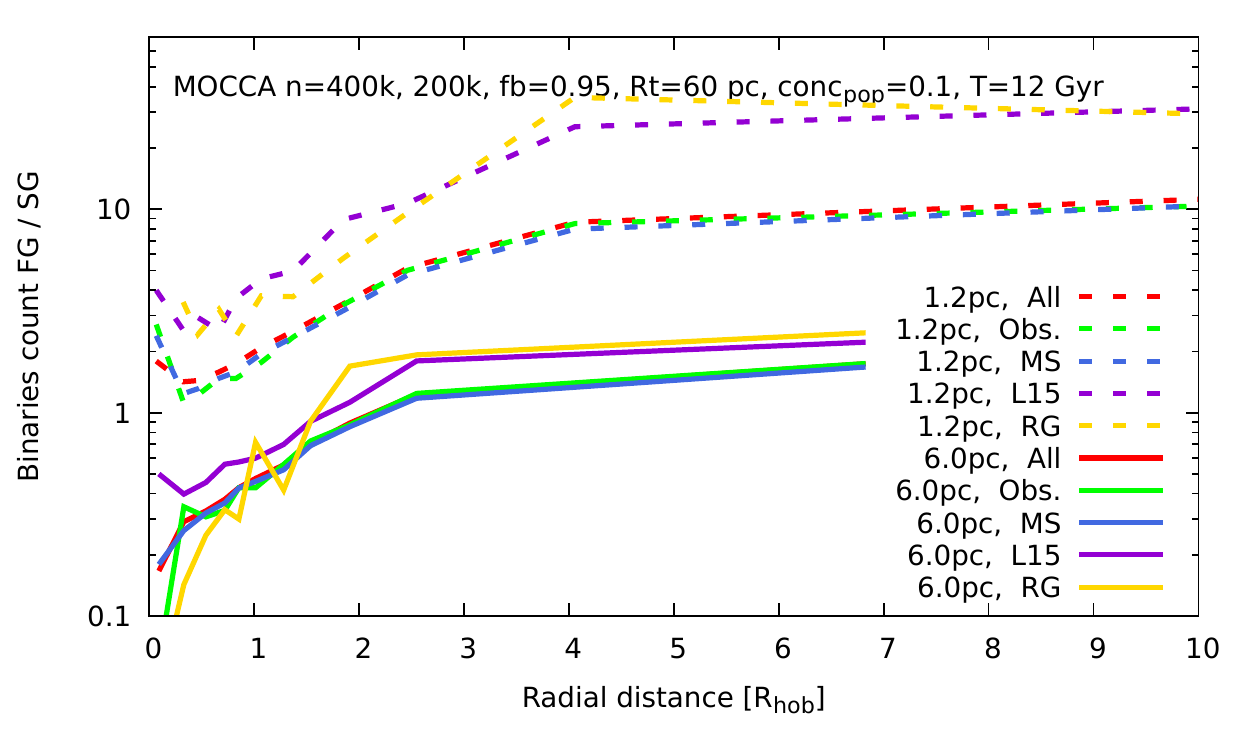}
  \caption{Binaries number computed in different ways for a set of \MOCCA simulations for two different Rh and for time 12 Gyr as a function of the radial distance scaled by \RHOB. Each line corresponds to one \MOCCA simulation which basic parameters are summarized in the caption. The simulations belong to two categories: underfilling cluster ($\rm R_h = 1.2$ pc, dashed lines), and filling cluster ($\rm R_h = 6.0$ pc, solid lines). The binaries are computed for the two groups in five different ways: all binaries, observational binaries, MS binaries, based on \citet[L15]{Lucatello2015}, and for red giants. For detailed definitions and a discussion see Section~\ref{s:DiffWays}.}
  \label{f:BinDiffWays}
\end{center}
\end{figure}


Figure~\ref{f:BinDiffWays} shows the ratios between FG and SG binaries as a function of the radial distance scaled by \RHOB, but computed in a few different ways than in Figure~\ref{f:Mix} for T~=~12~Gyr. There are two \MOCCA simulations presented on the plot. Their common initial properties are summarized in the caption. They differ only on the $\rm R_h$. The five first dashed curves on the plot concerns \MOCCA simulation with initial $\rm R_h = 1.2$~pc (small $\rm R_h$), which represents a class of underfilling models. In turn, the five bottom solid lines represent \MOCCA model with initial $\rm R_h = 6.0$~pc, which represents filling model (large $\rm R_h$). Each five lines represents (in that order) all binaries (of any masses), observational binaries, only MS binaries, binaries selected based on \citet[L15]{Lucatello2015}, and binaries with at least one red giant (RG). The observational binary is a binary if it fulfills the following conditions: 1. consists of two MS stars, 2. more massive star is $> 0.4~M_{\odot}$, 3. mass ratio $\rm q > 0.5$, 4. more massive star is less than $\rm M_{turnoff}$. The conditions roughly follows the prescriptions provided by \citet{Milone2012A&A...540A..16M}. The MS binary is simply a binary which consists of two MS stars of any mass. And finally, the criteria for L15 are following: $\rm m_1 > 0.7 M_{\odot}, log P/d > 2$, and $\rm q > 0.3$. These criteria have been chosen from the reported detection efficiency as a function of the orbital period reported by these authors \citep[Fig.~4]{Lucatello2015} and to impose the presence of a massive primary component.

Figure~\ref{f:BinDiffWays} shows that the ratio of binaries from different populations looks very similar for any method of computing the number of binaries if we take some subset of MS binaries. The shapes of the lines are the same as in Figure~\ref{fig:MixingOnly12Gyr}. For the models which are tidally underfilling ($\rm R_h = 1.2$ pc) the ratio has the clear low values in the central regions with the ratios above 1.0 (FG dominates). The higher values are in the outskirts of star clusters. In turn, for the tidally filling clusters ($\rm R_h = 6.0$ pc) the lowest values are also in the center but lower than 1.0, which means that SG is more numerous. At some distance this ratio gets larger than 1.0 and FG starts to dominate. The same features and the same physical processes are responsible for the shape of these curves as it was described in Section~\ref{s:Imprints} (e.g. Figure~\ref{fig:MixingOnly12Gyr}). Some differences appear for the cases of L15 and for RG binaries. For them the shape is the same, but the ratios are slightly larger, especially for underfilling model (yellow, and violet lines on Figure~\ref{f:BinDiffWays}). \citet{Lucatello2015} report binary fraction of $4.9\% \pm 1.3\%$ for FG and $1.2\% \pm 0.4\%$ for SG (with remark that the spectroscopic targets are located in a non-uniform way between $\rm 0.5 - 3 \times \rm R_{hob}$). We have computed them for models from Figure~\ref{f:ObsBinFrac} and the results, taken cautiously, are consistent within $2\sigma$ with all the above models, and within $1\sigma$ with the model tidally underfilling.

The tidally underfilling models give a better agreement with \citet{Lucatello2015} for the SG binary fractions (see Figure~\ref{f:BinDiffWays}). On the other hand, the tidally filling models in Figure~\ref{f:Rmax} suggest that they are better at obtaining the fraction of FG to SG stars from \citet{Milone2017MNRAS.464.3636M}. It mildly suggests that this discrepancy in binary fraction and overall fraction of SG stars may indicate that the SG stars are indeed born with a very high concentration, as suggested by \citet{Gratton2019A&ARv..27....8G} and \citet{Sollima2022}. This high concentration of SG stars (within an overall tidally-filling cluster) would result in increased destruction of SG binaries, while at the same time the overall fraction of SG stars would also increase as we will preferentially lose FG stars. We point out however, that \citet{Lucatello2015} results' are averaged over 6 GCs, and the number of examined stars is relatively small. Moreover, the average was done for the selected GCs for which \citet{Milone2017MNRAS.464.3636M} values of $\rm N_{FG} / N_{Tot}$ range from 0.175 (for NGC~104) up to 0.542 (for NGC~288). Thus, this conclusion is based on limited information and have to be treated cautiously. 

\begin{table}
    \caption{Number of FG binaries computed in a different way for one selected \MOCCA simulation (underfilling model with $\rm N_{1} = 400k$, $\rm N_{2} = 200k$, fb = 0.95, $\rm R_h = 1.2$ pc, $\rm R_t$ = 60 pc, $\rm R_{h SG} / R_{h FG} = 0.1$, T = 12 Gyr for FG -- the same as one of the models from Figure~\ref{f:BinDiffWays}). The first column is Lagrangian radius, the second column contains the number of all binaries (with any mass and types, in brackets is given the number single stars), and the next columns the number of observational binaries, MS binaries only (and number of MS single stars in brackets), and the number of binaries based on \citet[L15]{Lucatello2015}. The numbers of binaries are given for every Lagrangian radius separately, these values are not cumulative.}
    \centering
    \begin{tabular}{c|c|c|c|c}
    \hline\hline 
        Lagr. rad. & All bin. & Obs. bin. & MS bin. & L15\\
    \hline
        1\%  & 52 (181)       & 19   & 43 (122)    & 8\\
        10\% & 1695 (8303)    & 391  & 1230 (6226) & 186\\
        20\% & 2135 (11405)   & 496  & 1605 (8673) & 229\\
        30\% & 4174 (23125)   & 962  & 3289 (18371) & 350\\
        40\% & 4074 (23317)   & 813  & 3261 (19056) & 321\\
        50\% & 6499 (37835)   & 1210 & 5368 (31682) & 419\\
        60\% & 8181 (47927)   & 1411 & 6942 (41199) & 492\\
        70\% & 12055 (69087)  & 1929 & 10295 (60691) & 685\\
        80\% & 13206 (70470)  & 1953 & 11393 (62606) & 684\\
        90\% & 23352 (110171) & 3231 & 20264 (99045) & 1221\\
        100\% & 23100 (93287) & 3082 & 20015 (84252) & 1208\\
    \hline
    \end{tabular}
    \label{t:BinNumbers}
\end{table}

It is interesting to see that these different ways of computing binaries follow all of the binaries so well. From simulations one can have full information about binaries, their properties, number and spatial positions. However, for observations this is not possible. It is even more surprising if one take into account how many of the binaries are in these groups. The Table~\ref{t:BinNumbers} shows the number of binaries for one selected \MOCCA simulation (tidally underfilling model with $\rm R_h = 1.2$ pc from Figure~\ref{f:BinDiffWays}) for a number of Lagrangian radii. Additionally, for two columns the Table contains the number of single stars in equivalent categories for comparison. The table shows that the number of observational binaries are significantly less numerous than all binaries in the system. But yet, they follow the internal structure of the cluster in a very similar way, and thus they also follow the mixing between populations very well too. It is surprising to see this because the number of all binaries (all stellar types, all masses) for the model is still around 100k, but the number of e.g. observational binaries is around 14k only. Thus, the observational binaries should mimic the photometric observations and one can see that even such low number of observational objects can follow the global properties of the cluster. Moreover, the visible fraction of binaries varies between different regions of the cluster which additionally should disturb the picture: for Lagrangian radius 1\% there are 19 observational binaries from 52 all binaries which gives 36\%, whereas for the last Lagrangian radius only 3082 binaries are visible from all 23100 binaries, which gives only 13\%. However, even with such low numbers of binaries it is still possible to reflect the global properties of the whole cluster. It seems that observational binaries are sufficiently good proxy to probe the entire cluster. This result is probably connected with the fact that at late evolution times there is only a small difference between masses of all stars/binaries and also the number of non MS binaries is relatively small. 

\begin{figure}
\begin{center}
    \includegraphics[width=0.99\linewidth]{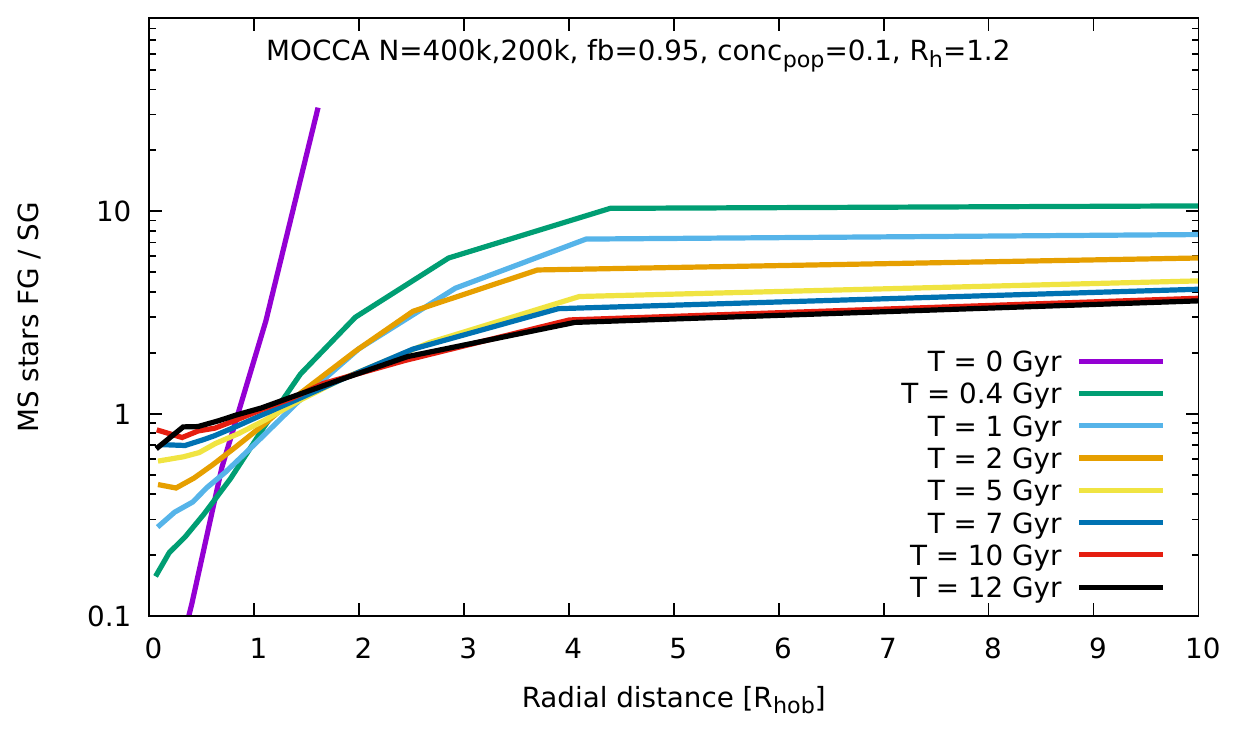}
    \includegraphics[width=0.99\linewidth]{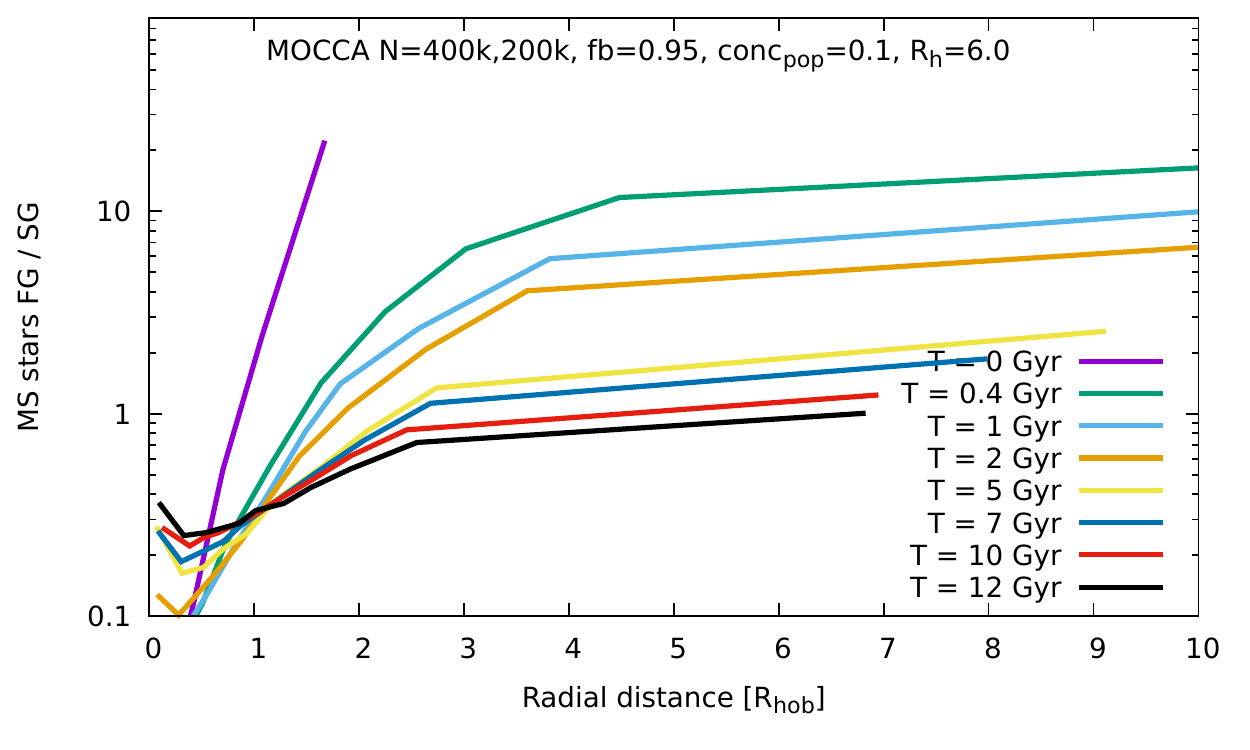}
  \caption{Description as in Figure~\ref{f:MixDiffRh} and also with the
  same \MOCCA models. The only difference is that only two 
  \MOCCA models are shown here (for smallest and largest $\rm R_{h}$)
  and plots are prepared based on MS single stars 
  only (not binaries). For details see Section~\ref{s:DiffWays}.
  }
  \label{f:MixMS}
\end{center}
\end{figure}

And finally, Figure~\ref{f:MixMS}, presents two \MOCCA simulations -- the same as from Figure~\ref{f:MixDiffRh} but only with two initial $\rm R_h = 1.2, 6.0$ and prepared with MS single stars (not binaries). The plot shows that even taking only MS single stars into account (not binaries), to track the mixing between the two populations, seems to be enough. MS single stars present the same features as the binaries described in details in Section~\ref{s:Imprints}. The only noticeable difference is that for MS single stars the ratios between FG and SG seem to have values consequently closer to 1.0. In Figure~\ref{f:MixDiffRh} for \MOCCA simulation with $\rm R_h = 1.2$ the ratio between FG and SG binaries below \RHOB has the value about 2.0, whereas for MS single stars in Figure~\ref{f:MixMS} this ratio is about 1.0. In turn, for \MOCCA model with $\rm R_h = 6.0$ it is the ratio with value about 0.2 for binaries, and for MS stars about the value 0.3. We did not investigate these differences closely. From now on in this paper FG and SG mean both FG, SG binaries, or FG, SG stars (unless specified otherwise).

\subsection{Binary fractions}
\label{s:BinaryFractions}

\begin{figure}
\begin{center}
  \includegraphics[width=0.99\linewidth]{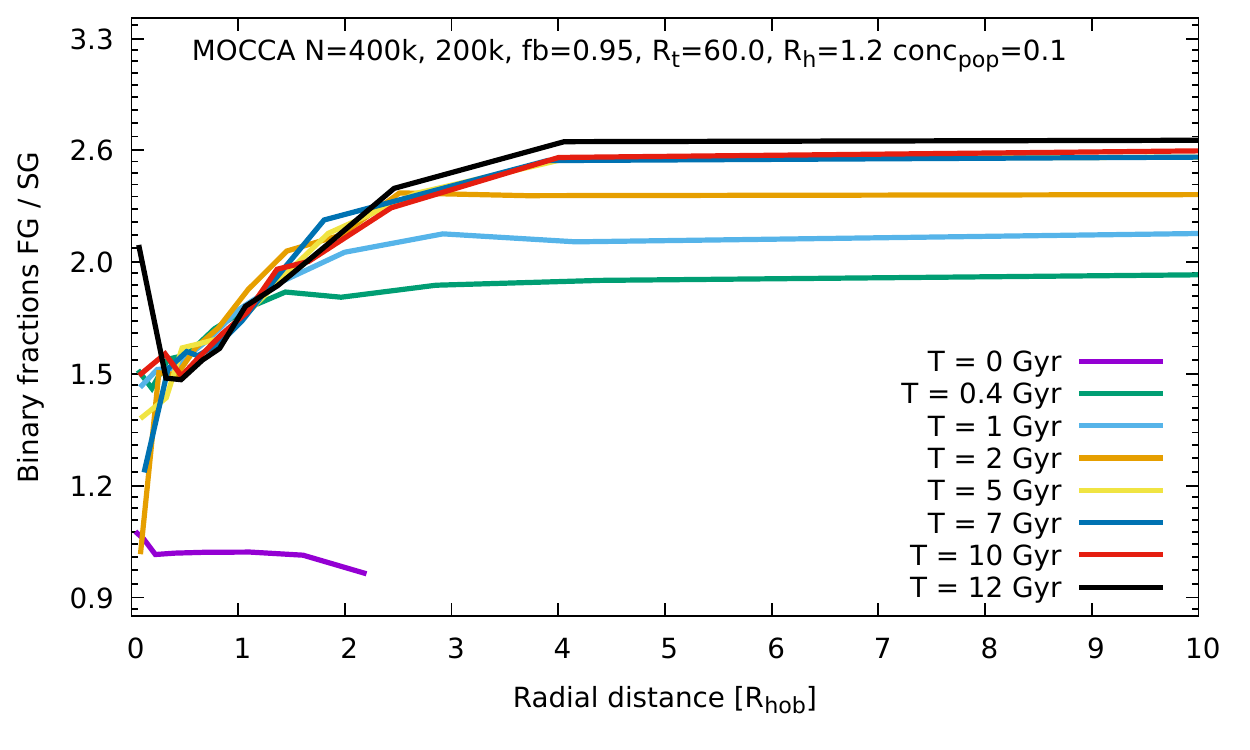}
  \caption{Radial profile of ratios between binary fractions from FG and
  SG for a few selected times for one \MOCCA simulation with 
  concentration parameter between populations equals to 0.1 (\CONCPOP = 0.1) as a function of radial distance scaled by \RHOB. 
  For details see Section~\ref{s:DiffWays}}
  \label{f:BinFrac}
\end{center}
\end{figure}

The same trend as in Figure~\ref{f:Mix} for binaries count one can obtained also with the binary fractions. A binary fraction is defined here simply as a ratio between the number of binaries to the number of binaries and single stars (for any given population). The Figure~\ref{f:BinFrac} shows radial profile of ratios between binary fractions from FG and SG for a few selected time-steps for one \MOCCA simulation. The concentration parameter \CONCPOP between populations equals to 0.1. As expected, the initial binary ratio is equal to 1.0 for the entire star cluster (violet line in Figure~\ref{f:BinFrac}). Later with time the binary ratio between FG and SG increases, and eventually after Hubble time there is significantly more binaries from FG than the SG in the central regions of the cluster -- the same evolution of populations is observed with the larger binary fraction from FG and SG visible on the Figure~\ref{f:Mix}, or Figure~\ref{f:MixDiffRh}. It is also interesting too see that the ratio increases quickly over about first 2 Gyr, and after that the ratio increases much more slowly. This is because for underfilling models for substantial part of their evolution they behave as isolated clusters -- they nearly freely expand. FG was initially more extended than SG, so it expands a bit faster. Additionally, because of increase of \RHOB the same value on X-axis means actually larger physical distance. SG binaries needs more time to access the same physical distance as FG binaries. This leads to increase of the ratio between FG and SG binaries. Later, when cluster becomes tidally filling the described process slows down and SG binaries easier can catch FG binaries. 

\begin{figure}
\begin{center}
  \includegraphics[width=1.0\linewidth]{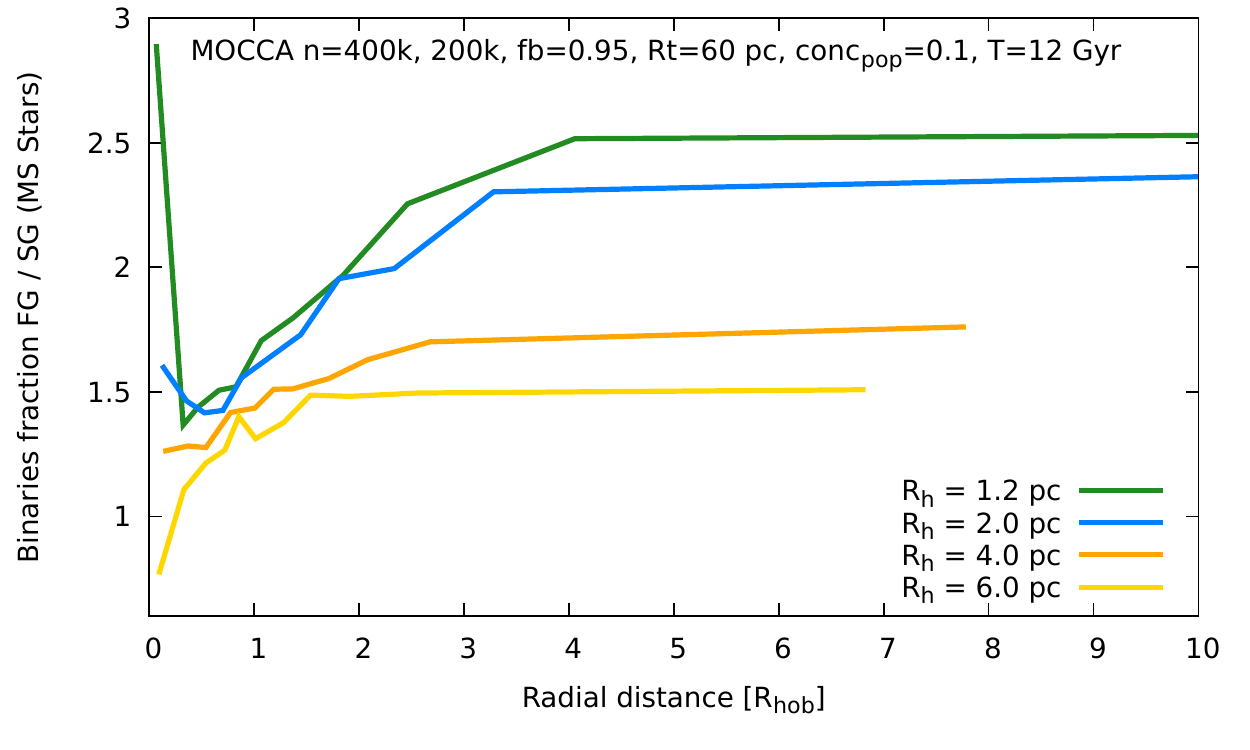}
  \caption{Binary fractions between FG and SG (only MS stars) for a set of \MOCCA simulations which differ in half-mass radii (from 1.2 pc -- tidally filling, up to 6.0 pc -- filling models) as a function of radial distance scaled by \RHOB. The rest of the initial parameters are summarized in the caption. The binary fractions between FG and SG show the same evolution of the mixing between two populations as computed from the number of all binaries (see Figure~\ref{f:BinFrac}, or \ref{f:BinDiffWays}).}
  \label{f:ObsBinFrac}
\end{center}
\end{figure}

Similar evolution of the binary fraction ratios stands valid also for e.g. MS binaries will be taken into account only. Figure~\ref{f:ObsBinFrac} shows the ratio between binary fractions FG and SG for a few selected \MOCCA simulations which have the same initial conditions (given in the caption), but have different $\rm R_h$. The ratios between the binary fractions of FG and SG are computed here using only MS binaries and stars. The models with small $\rm R_h$ are tidally underfilling, whereas the one with large $\rm R_h$ values are filling. One can see that the evolution of the ratio between two populations is basically the same as it was already discussed in Section~\ref{s:Imprints}. The ratio between the binaries fractions have a peak in the center of the cluster for the tidally filling models (small $\rm R_h = 1.2$ pc) which is caused by quickly burned out SG and as a result FG starts to dominate (the same evolution like black curve in Figure~\ref{f:BinFrac}). In turn, for tidally filling models, there the ratio between the binary fractions is smaller than 1.0 for central regions (SG dominates), and at some distance (around \RHOB) the ratio gets larger than 1.0. It means that FG dominates, and the ratio raises to the value around 1.5 which is very close to the initial one.

Figure~\ref{f:ObsBinFrac} shows also that the ratio between binary fractions FG to SG is smaller than 1.0 only in the very center of a cluster (much below \RHOB) for tidally filling clusters (large initial $\rm R_h$ values). Only there SG binary fractions are larger than for FG. It seems that SG for tidally filling clusters are not dissolved so efficiently like for much more dense tidally underfilling clusters. This could be potentially important imprint in observational data which could help to narrow down the initial conditions for star clusters.

\begin{figure}
\begin{center}
  \includegraphics[width=1.0\linewidth]{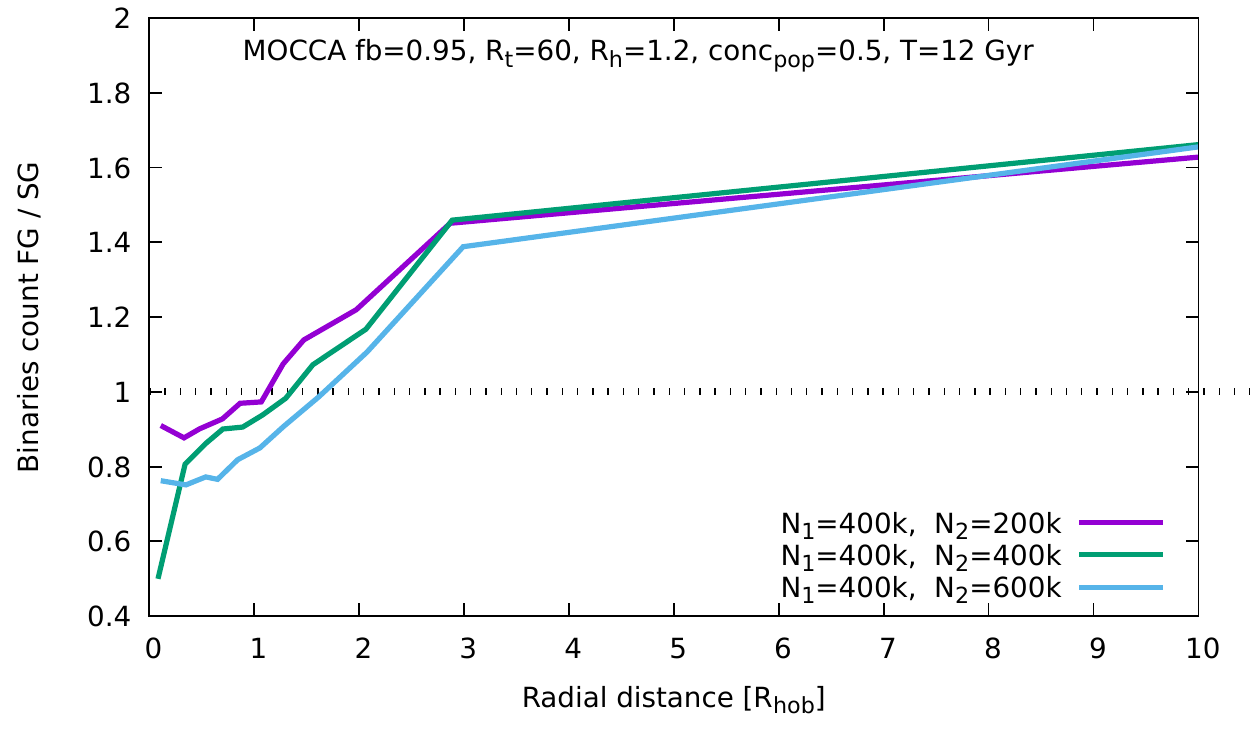}
  \caption{Ratio between binaries number of FG to SG for three \MOCCA simulations for 12 Gyr as a function of radial distance scaled by \RHOB. All \MOCCA simulations have fb = 0.95, $\rm R_t = 60$, $\rm R_h = 1.2$, \CONCPOP = 0.5 which means that SG is two times more concentrated than FG. They differ in the initial number of second population ($\rm N_2 = 200k, 400k, 600k$), thus their initial mass differs too. The ratios are scaled by the initial number of binaries for easier comparison. For details see Section~\ref{s:DiffWays}.
  }
  \label{f:BinCountDiffMass}
\end{center}
\end{figure}

The initial mass of the model does not change the ratios of binaries number of FG to SG too. Figure~\ref{f:BinCountDiffMass} presents three \MOCCA simulations (fb = 0.95, $\rm R_t = 60$, $\rm R_h = 1.2$, \CONCPOP = 0.5) at 12 Gyr as a function of radial distance scaled by \RHOB. They have different only the initial number of SG stars ($\rm N_2 = 200k, 400k, 600k$). The ratios are scaled by the initial number of binaries in order to compare them on one plot. As a result the models have different initial masses. The plot shows that the evolution of the ratio between FG and SG binaries is not affected much by the initial mass. Only the very center of the cluster shows larger differences between different models, but this is most likely due to fluctuations of small number of binaries. This can have rather important implications for the observations because it shows that the mixing does not depend much on the initial mass of the star cluster. It seems that the initial degree of tidal filling or underfilling matters more.

\section{Discussion}
\label{sec:Discussion}

The analysis of the \SURVEYTWO simulations revealed a number of interesting features of multiple stellar populations which could be helpful to understand some of the observational properties of the real GCs. In this section we will discuss the results of our analysis especially with respect to the boundaries of the initial conditions of GCs. We would like to remind that we are working in a scenario in which SG was formed from a FG ejecta and pristine gas in the central regions of the FG cluster. We also focus in this work mainly on the \MOCCA simulations with $\rm N_{FG}$ = 400k, and $\rm N_{SG}$ = 200k (see Table~\ref{tab:MoccaSurvey2}) because SG as it is expected from theory and observations is less numerous than FG and this allows to compare to some extent \MOCCA simulations with observations. 

\begin{figure}
\begin{center}
  \includegraphics[width=1.0\linewidth]{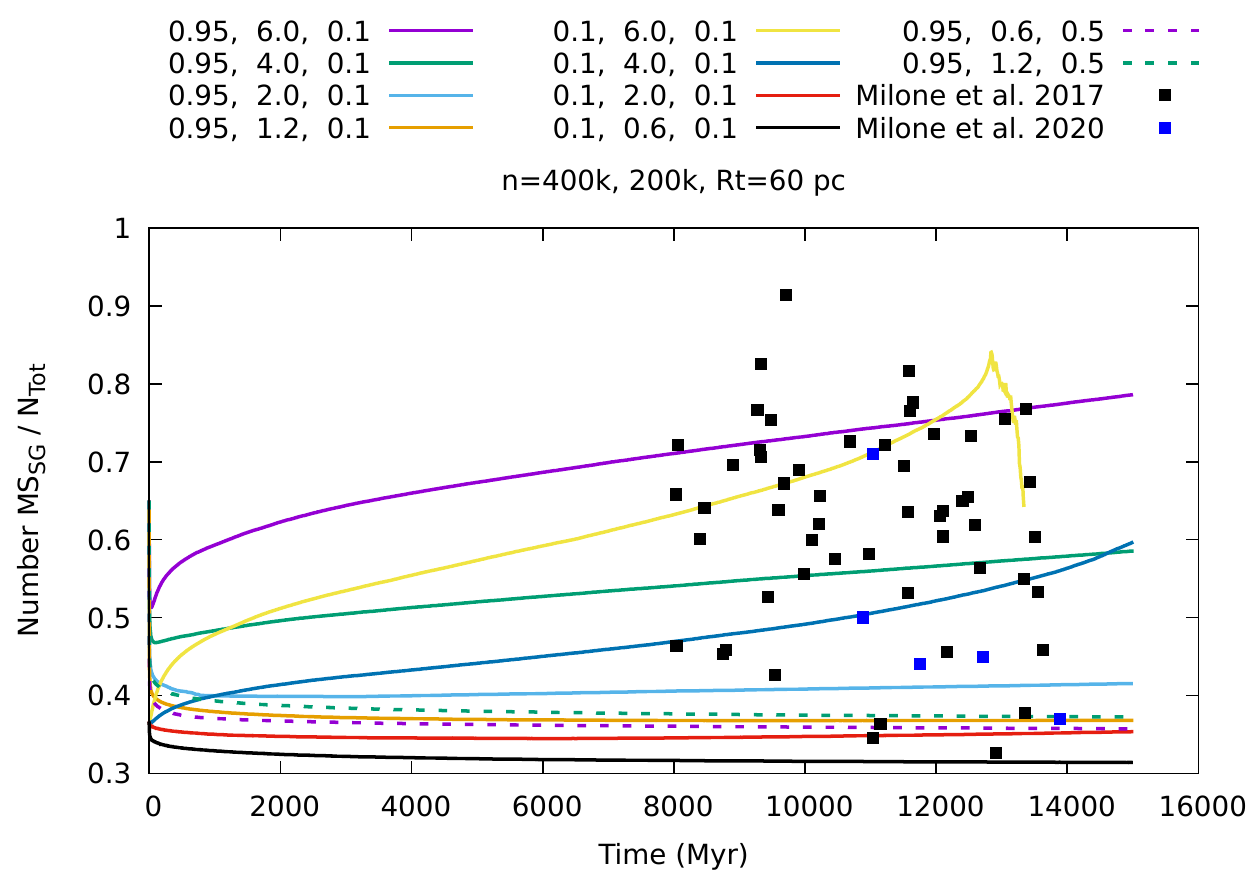}
  \caption{Number of MS SG as a fraction of all objects for a number
  of selected \MOCCA simulations presented with lines as a function 
  of time. The black and blue dots represent values for real Milky Way GCs 
  \citep{Milone2017MNRAS.469..800M,Milone2020MNRAS.492.5457M}
  and are spread in time randomly between 8 and 14 Gyr for clarity. 
  The \MOCCA simulations 
  are characterized with three values: binary fraction, 
  half-mass radius, and \CONCPOP respectively. They represent
  models which are tidally underfilling and strongly
  tidally filling. }
  \label{f:NSGNtot}
\end{center}
\end{figure}

We emphasize that the primary goal of this study was to explore the mixing process of FG and SG binary stars and the range of models explored was not meant to provide a comprehensive survey of initial conditions mimicking the Milky Way GCs. However, as it turned out, the models cover the properties of the Galaxy GCs surprisingly well (see Figure~\ref{fig:Holger}). Figure~\ref{f:NSGNtot} shows MS SG as a fraction of the total number of objects for a number of \MOCCA simulations (as lines) and for real Milky Way GCs \citep{Milone2017MNRAS.469..800M} as black dots. Each \MOCCA simulation (400k FG, 200k SG, $\rm R_t = 60$ pc) is summarized with caption consisting of three values: binary fraction, half-mass radius and \CONCPOP. One can see that \SURVEYTWO models cover surprisingly well the real GCs population in terms of the fraction of SG stars of total number of objects. The models of \SURVEYTWO analyzed in this paper can reproduce the whole spectrum of MS~SG/NTot fractions -- the ones for which SG dominates and also the ones with FG more numerous. The models can reproduce basically any value from the observations (from $\sim 0.3$ up to $\sim 0.8$, the point $\sim 0.9$ concerns $\rm \omega Cen$ which might not be a good representative of GCs) despite the fact that they differ in $\rm R_h$ and \CONCPOP initial values only (we already described in Section~\ref{s:Imprints} that initial binary fraction does not influence much the results). It is quite unexpected finding, but gives high confidence that the results of \SURVEYTWO can be compared with the real observations. Even with our limited variety of initial conditions (e.g. limited initial mass range for models) we cover MW GCs well, and thus we can discuss all of the observational features presented in the literature.

\SURVEYTWO models can well reproduce fractions of SG stars for the known range of observed values in the real GCs. The main thing which seems to be varying the models from \SURVEYTWO from the point of view of these fractions is the tidal filling of the cluster. Figure~\ref{f:NSGNtot} shows a clear division of models into two groups: the ones for which number of SG increases (tidally filling models), and the ones for which it decreases (tidally underfilling). In the case of \SURVEYTWO models this division is around $\rm R_h = 2.0$ pc. Among GCs which have the lowest SG stars there is e.g. M~4 (NGC~6838) with $\sim 35\%$ of SG with respect to all MS stars \citep[Table~1]{Milone2020MNRAS.492.5457M}. The \MOCCA models which are consistent with low SG fractions are the models which are tidally underfilling (low $\rm R_h$ values in Figure~\ref{f:NSGNtot}). One can see that the initial SG fraction drops for them significantly in the beginning of cluster evolution due to fast mass loss, which expands the cluster (due to high densities many dynamical interactions eject a lot of objects). Then the ratio drops further and settle around the values $0.3 - 0.4$. These are the models which are initially very dense, go to a core collapse very quickly and after the collapse they expand roughly homogeneously. Moreover, SG binaries burns out very quickly because it is initially very dense, the binaries are being destroyed or ejected more than for FG. In turn, the GCs which have the highest values of SG are e.g. 47~Tuc (NGC~104) or M13 (NGC~6205) which have $\sim 80$\% of SG stars \citep[Table~1]{Milone2020MNRAS.492.5457M}. These are old, still massive and large GCs. Their SG fraction is consistent with \MOCCA models which were initially tidally filling, and thus also not very dense clusters (high $\rm R_h$ values in Figure~\ref{f:NSGNtot}). These are the models in which $\rm R_{h FG}$ is large and thus many of FG stars are close to the tidal radius. This in turn causes that the models lose preferentially FG stars, and thus leaving the cluster eventually with overpopulated SG stars. This is only a hint about the initial GC conditions. To get better constraints and comparison with MW GCs we will need to check also other cluster properties which can be different than the ones following the present \SURVEYTWO models.

The results of \SURVEYTWO simulations work within the scenario in which it is assumed that SG was formed from a FG ejecta and pristine gas in the central regions of the FG cluster. In that scenario it is expected that SG should be concentrated in the center of the cluster and should have higher densities than FG stars. These are the models which initially have SG less numerous than FG and the \CONCPOP values are below 1.0 (SG is more concentrated than FG). A number of such models are shown as lines in Figure~\ref{f:NSGNtot}. It is important to point out that starting with such models we are able to reproduce all cases when it comes to the fraction of SG with respect to total number of stars. 
\MOCCA simulations provide some interesting implications for the so-called mass-budget problem. In the literature there are known GCs for which SG is more numerous (see Section~\ref{sec:Intro}). This creates a challenge for interpretation because it is hard to expect that the formation of multiple population would create more SG than FG stars. Up to now one of the most controversial implications is that GCs should be much more massive during formation \citep{Ventura2014}. Then, GC has to lose a lot of FG stars, which would contribute to host galaxy formation, and also they would provide some contribution to the reionization of the Universe (e.g. \citealt{Renzini2015}). We find from \SURVEYTWO models that the main parameter which matters in that respect is how the initial models is tidally filling its potential with respect to the host galaxy and also the concentration parameter. For the models which are initially tidally filling there is basically no problem to have at 12 Gyr for SG to dominate the cluster in terms of the number of stars. It is in agreement with the results of previous study from \citet{Vesperini2021MNRAS.502.4290V} and \citet{Sollima2022}. Figure~\ref{f:NSGNtot} shows how numerous are MW GCs which have SG larger than FG. It seems that these are the clusters which were formed close to be tidally filling. In order to have stronger claims about the mass-budged problem we need more models and we need to check if other GCs observational parameters are also recovered. However, our models give some support for other authors claiming that the mass budged can be overcame.

The potential presence of an IMBH \citep{Baumgardt2004ApJ...613.1143B, Trenti2007MNRAS.374..857T} does not seem to influence the findings of this paper (for details of the IMBH formation in \MOCCA simulations in \SURVEYONE see \citealt{Giersz2015MNRAS.454.3150G}). The IMBH influences the dynamical interactions, and the core properties of star clusters, but yet it seems not to have any significant influence the fractions of FG or SG populations. There are models in \SURVEYTWO which initially are tidally filling, or tidally underfilling (e.g. see Figure~\ref{f:MixRhConst}) and which produced (or not produce) a massive IMBH, but the evolution of the fraction of the population seems to be unaffected by an IMBH. Moreover, we observe in some of the models with an IMBH, the homogeneous expansion for the most dense clusters (Lagrangian radii increase at a steady pace with time). In such models the IMBH works like an energy source. After core collapse it behaves like a set of binaries in the center of a cluster and prevents further collapse of the core. After that we also observe the homogeneous expansion of the cluster described in the Section~\ref{s:Imprints}. We plan to investigate IMBH influence on the multiple population more deeply in the next papers.


\begin{figure}
\begin{center}
  \includegraphics[width=1.0\linewidth]{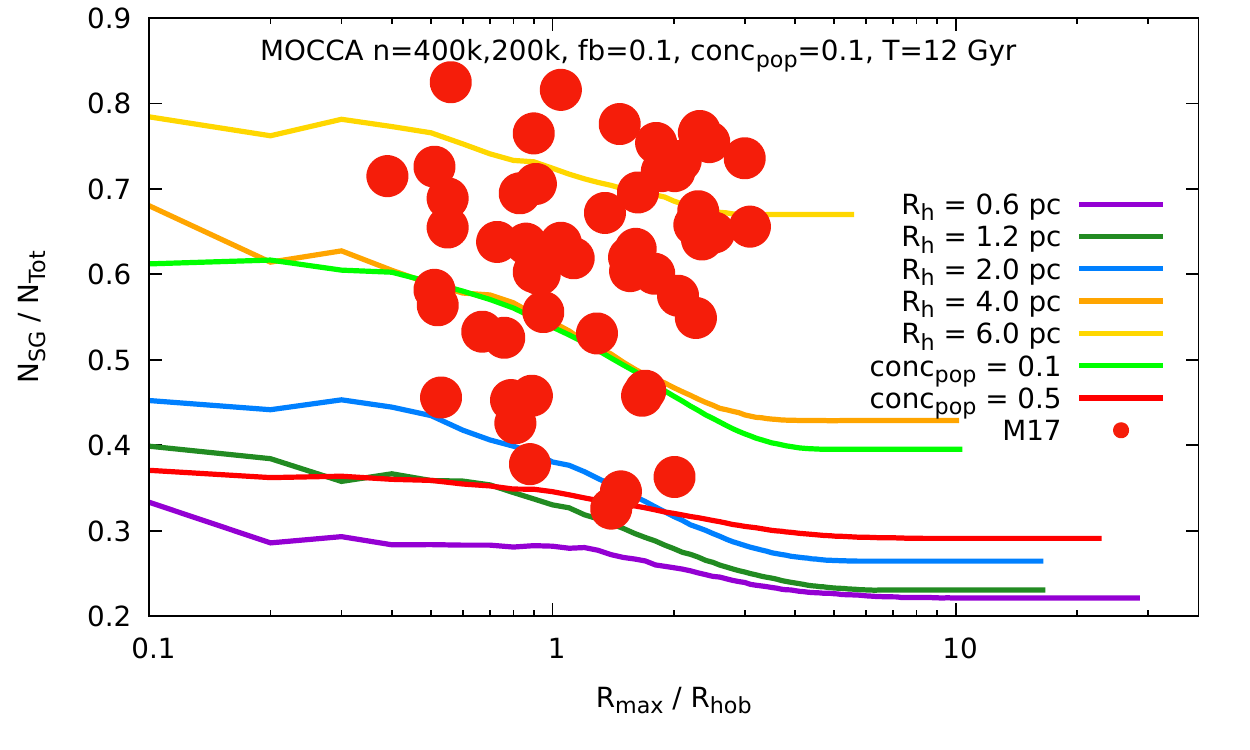}
  \caption{Fractions of SG of the total number of stars ($\rm N_{SG} / N_{Tot}$) for Milky Way GCs \citep[Table~2, abbreviation M17]{Milone2017MNRAS.464.3636M} as a function of the maximum radius of the observations normalized to the half-light radius ($\rm R_{max} / R_{hob}$). Additionally, on the plot there is a few \MOCCA simulations with the fractions of SG stars as a function of the distance from the cluster center in the units of half-light radius. Five first \MOCCA simulations have  common initial properties, summarized in the caption, and different only initial half-mass radii. There are two additional \MOCCA simulations: $\rm conc_{pop} = 0.1$ with $\rm R_h = 4.0$ pc, fb = 0.95, and $\rm conc_{pop} = 0.5$ with $\rm R_h = 1.2$ pc, fb = 0.95 given for comparison. Different $\rm R_h$ represent star clusters initially tidally underfilling (e.g. $\rm R_h = 0.6$ pc), and tidally filling (e.g. $\rm R_h = 6.0$ pc). The data present FG fractions only for 12 Gyr. For details see text.}
  \label{f:Rmax}
\end{center}
\end{figure}

Fractions of SG with respect to the overall number of stars can vary significantly in different radial parts of a star cluster (e.g. Figure~\ref{f:ObsBinFrac}). We have checked how the fraction of SG changes from the central parts of a cluster to the outskirts and compared a few \MOCCA simulations with available observations. Figures~\ref{f:Rmax} shows the fractions of SG for Milky Way GCs \citep[Table~2]{Milone2017MNRAS.464.3636M}. For comparison, the figure shows a set of \MOCCA simulations which have the initial parameters summarized in the caption and only different initial half-mass radii. 
The \MOCCA simulations show SG fractions only for 12 Gyr for a number of Lagrangian radii scaled by \RHOB. Figure~\ref{f:Rmax} shows that the values from \SURVEYTWO cover the real observations very well. \SURVEYTWO is able in principle to reproduce the whole observed range from observations. The fraction of SG depending on the maximum distance of observations from the center of the cluster ($\rm R_{max}$) show a very distinct evolution for different initial concentrations with respect to the tidal radius. The first group have lower changes of the fraction of SG with respect to $\rm R_{max}$. Violet curve for tidally underfilling cluster ($\rm R_h = 0.6$ pc) have in the center SG fraction $\sim 0.3$, and in the outskirts of the cluster value around $\sim 0.25$. This is a difference of the order of 0.05. In turn, for the clusters which started initially tidally filling the SG fraction for low $\rm R_{max}$ is high (SG dominates strongly), then with larger $\rm R_{max}$ the fraction drops more significantly. Eventually for the model with $\rm R_h = 6.0$ pc it drops below 0.7 ($> 0.1$ difference), and for the model with $\rm R_h = 4.0$ pc the difference is even larger (from 0.7, to 0.45). For tidally filling cluster the ratio drops more significantly with $\rm R_{max}$. It would be very informative to compute $\rm N_{SG} / N_{Tot}$ values for MW GCs for a number of radii up to $\rm R_{max}$ and then compare them with the simulations. The fraction of SG stars, if computed for different values of $\rm R_{max}$, could be an interesting tool for probing the initial conditions of the clusters. 

Interestingly, MSP distribution may also help to narrow down the initial conditions for GCs. Figure~\ref{f:Rmax} shows that the models from \SURVEYTWO with initial $\rm R_h = 0.6$ pc seem to have initial conditions too dense. The models with such densities do not cover any of the known MW GCs. This is very helpful e.g. in determining the initial conditions for the next set of \MOCCA simulations.

The work in this paper is a natural extension of the work done in \citet{Vesperini2021MNRAS.502.4290V} -- but e.g. with the most up-to-date version of the \MOCCA code and for many more models. In \citet{Vesperini2021MNRAS.502.4290V} it was found that in order to have $\rm M_{SG}/M_{tot}$ values observed $\sim 0.35 - 0.9$ one has to lose preferentially FG stars. This occurs during the first Gyr when system expands and responds to the mass loss of FG which is less concentrated. In our study we connect this with by how much the initial model is settled with respect to the tidal radius. Moreover, in \citet{Vesperini2021MNRAS.502.4290V} the final values for $\rm M_{SG}/M_{tot}$ are between 0.53 and 0.8. In this work we are also able to reproduce lower values -- characteristic for underfilling clusters. 

The results are also consistent with \citet{Sollima2021MNRAS.502.1974S} who find that SG observed today depends uniquely on the ratio between the initial cluster mass and the half-mass radius of SG. The efficiency of SG being destroyed decreases with time after the cluster initial mass loss. Thus the SG fraction observed today is strongly sensitive to the initial size of the SG. They use this relation to constrain the initial conditions of GCs and concluded (with a few assumptions) that SG must have formed with a typical half-mass radius smaller than 0.5--1 pc. In turn, FG is more extended and more exposed to the tidal field during the cluster evolution. They find that the final FG binary fraction depends on the strength of the tidal field which determines the variation of cluster mass and size. We find the same conclusions but in our models we see it through how much the models are filling or underfilling -- this determines the final fractions of SG as we already described in previous paragraphs. 

\section{Conclusions}
\label{s:Conclusions}

The main purpose of the paper was to test mixing in dense GCs for various initial conditions for two populations of stars (FG and SG) for real size GCs -- the goal difficult to obtain currently with direct \textit{N}-body codes. We work within scenario by \citet{Calura2019MNRAS.489.3269C}, in which SG forms in the very cluster center, after some time delay (usually a few dozens of Myr), from gas lost due to stellar winds of AGB stars and pristine gas reaccreated by GC during its movement through gas cloud left after formation (some small deviations are described in Section~\ref{sec:InitialConditions}). The main conclusion of the paper concerns how initially tidally filling or underfilling clusters influence the relative number of FG to SG stars. The results of this study are the following:

(i) We find that the most important factor affecting the relative fraction of FG or SG is the initial tidal filling of a cluster. In order to have FG more numerous than SG, a star cluster has to be tidally underfilling since the cluster can expand without losing FG stars during the initial expansion phase. On the other hand, tidally filling clusters tend to have more SG stars than FG due to the preferential loss of FG stars through the tidal radius in accordance with their initial position.
Binaries also have the same trend, tidally filling clusters have more SG binaries and underfilling clusters have more FG binaries (Figure~\ref{fig:MixingOnly12Gyr}). But the differential evolution of binaries between FG and SG is mostly driven by the dynamical interactions at the central regions. The binaries from SG are much more efficiently destroyed or ejected due to the dynamical interactions at the dense central regions, especially in a underfilling cluster (see Section~\ref{s:Imprints}). In turn, to have a more numerous SG binaries, a star cluster has to be initially tidally filling system whose central density is not very high to destroy or kick SG binaries efficiently but FG binaries can escape from the cluster easily.

(ii) Even though the \SURVEYTWO models presented in this paper were not designed to mimic Milky Way GCs, their cover observational parameters very well (see Section~\ref{sec:InitialConditions}). Our models can reproduce the observed range of fractions of SG stars with respect to the total number of stars for known values of MW GCs (see Figure~\ref{f:NSGNtot}). We find that the MW GCs which have the lowest values of SG to total number of stars (e.g. NGC~6496 with 33\% of SG; see \citealt[Table~2]{Milone2017MNRAS.469..800M}) are consistent with our \MOCCA models of tidally underfilling cases ($\rm R_h$ = 0.6~pc in Figure~\ref{f:NSGNtot}). In turn, MW GCs with the highest SG fractions (e.g. 47 Tuc, M13 which has 80\% of SG stars) are consistent with \MOCCA models initially tidally filling. They are old, still massive and large GCs. For these models $\rm R_{hFG}$ are large and thus many of FG stars are closer to the tidal radius. Thus, these models lose preferentially FG stars, and as a result, SG dominates at the present-day, which is consistent with the conclusions of \citet{Vesperini2021MNRAS.502.4290V}. We note also, that \SURVEYTWO have a minimum initial number of SG/FG of 0.5, and it is not clear whether very large final fractions of SG/FG could be reproduced assuming smaller initial fractions \citep{Bastian2015}.
This has also an implication to so-called mass-budget problem. MW GCs with more numerous SG create a challenge for the interpretation because it is hard to explain why SG can be more abundant. Some scenarios \citep{Ventura2014,Renzini2015} tried to interpret the mass-budget problem, but there has been no definitive proof yet. Our \MOCCA simulation results can shed a light on solving the mass-budget problem by reproducing any fraction of SG value of the real MW GCs without any unrealistic assumptions or postulates. However, we need to run more models, and check if other parameters of our models are in good agreement with MW GCs properties. We are aware that we get good agreement with respect to one parameter but we will have to check also another ones in order to make more conclusive statements.

(iii) we examined the fractions of SG with respect to the overall number of stars (Figure~\ref{f:Rmax}) as a function of maximum radial distance ($\rm R_{max}$) for which they were measured \citep[Table~2][]{Milone2017MNRAS.464.3636M}. \MOCCA models can reproduce the whole range of values from observations, but we find that $\rm R_{max}$ can vary significantly within one GC if we take $\rm R_{max}$ to only a fraction of $\rm R_{h}$ or a few half-mass radii. For e.g. tidally underfilling models the fraction $\rm N_{SG} / N_{Tot}$ varies from $\sim 0.7$ in the very center to about $\sim 0.45$ in the outskirts (see details in Section~\ref{sec:Discussion}). 

(iv) We find also that the ratio between binary count from FG to SG which takes into account only MS stars is a very good proxy for the entire cluster. It provides the same conclusions for mixing as those coming from all binaries, or only observational ones (see Section~\ref{s:Imprints}). The same behavior of the ratio between two populations let us believe that even by observing a small fraction of visible MS binaries one can actually probe the whole star cluster (i.e. all binaries) at least from the point of view of mixing populations.


(v) The conclusions of this paper are consistent between the models with the initial binary fraction 10\%, and 95\%. It is known, that for the latter case many binaries are weak from the point of view of the mean binding energy of binaries and many of them are quickly destroyed. But still, the number of primordial binaries has a profound effect on the star cluster evolution. It is interesting to see that the initial binary fraction does not affect much the ratio between number of binaries from two populations. 


(vi) Preliminary examination indicate that a presence of IMBH in some models do not affect the conclusions of this work. The IMBHs, even with masses $\rm> 1000.0~M_{\odot}$ do not affect significantly the ratio between binary fractions of two populations. From this point of view, any try to infer constrains on the initial star clusters concentrations should be safe too.


In this paper we have concentrated on the number of FG and SG stars through the prism of tidally filling or underfilling clusters. There is more parameters which differ simulations within \SURVEYTWO simulations. We plan to explore their influence on the FG and SG populations in follow up papers (e.g. IMBH and BH subsystem influence, time delay between FG and SG formation). We did not check e.g. whether different $\rm M_{max}$ values (Table~\ref{tab:MoccaSurvey2}) have any influence on the FG, SG populations or their relative number.


\section*{Acknowledgments}


We would like to thank Enrico Vesperini for very insightful comments and suggestions which helped to significantly improve this paper. We thank also the referee for all the comments and suggestions, which improved the paper even more.

This research has been partly financed by the Polish National Science Centre grants: 2016/23/B/ST9/02732, 2018/30/A/ST9/00050, 2019/32/C/ST9/00577, 2021/41/B/ST9/01191. AA acknowledges support from the Swedish Research Council through the grant 2017-04217.


\section*{Software}

\MOCCA code is open source\footnote{\url{https://moccacode.net/license/}} for our collaborators. We are open to start new projects, in which one could use already existing \MOCCA simulations, or start new numerical simulations.

\textsc{beans}\footnote{\url{https://beanscode.net/download/}} software is open source and it is freely available for anyone.





\section*{Data availability}
The data underlying this article will be shared on reasonable request to the corresponding author.

\bibliographystyle{mn2e} 

\bibliography{biblio}



\appendix

\section{Output files}
\label{s:AppendixA}

The output files produced for every \MOCCA simulation consists of around 20 files. They were designed to store global properties of star clusters, profiles, snapshots and various events which occur for and between stars (e.g. stellar evolution, dynamical interactions, merger). The density of the stored information depends on the type of the information. Global star cluster parameters are computed for every time-step, snapshots are written by default every 50~Myrs and stellar events (e.g. mass loss, dynamical interactions, natal kicks properties) are all written to the output files. The overall density for the output was designed to be as compact as possible but providing as much information as possible for the versatile needs of any user.

An additional explanation is needed on how different interactions are treated for some multiple populations. For mass transfers we consider marking a star as a mixed population if the star gained some mass larger than $1.0e^{-7} M_{\odot}$ (an arbitrary chosen value). The star which lost the mass keeps its population id. When it comes to dynamical collisions, if two stars are coming from two different populations, we always mark the star as a mixed population.

A mixed population id for a star is designed in such a way that it holds compact and partial information about the history of mixing between different populations. In other words, the mixed population id is larger for a star for which the mixing (mass transfers or dynamical collisions) happened more times. The mixed population id is a concatenation of two populations ids of interacting stars and it is done according to the following rules:

\begin{itemize}
	\item pop1+pop2 $\rightarrow$ (01)(01) $\rightarrow$ gives the id 101 (first two digits
      in the right brackets tells us how many merger stars come from
      population 1 and two next digits from the next brackets -- how many stars come from
      the population 2). In this case there is one star from population 1 and
      one from population 2.

    \item  pop2+101 $\rightarrow$ (01+01)(01) $\rightarrow$ number 201 - in this
    case there are
      in total two stars (or mass transfers) from population 2 and one star from population 1.

    \item  pop1+201 $\rightarrow$ (02)(01+01) $\rightarrow$ number 202 - in this
    case there are in total two stars (or mass transfers) from population 2 and two
    stars from population 1.

	\item further population ids are created accordingly.

\end{itemize}

The following scheme is shown for simulation consisting of two populations. For the simulations started with more simulations there is only a larger number of significant digits (e.g. (02)(01)(03) $\rightarrow$ two stars or mass transfers from population~3, one from population~2, and three from population~1). This mixed population id is designed only to help to see how complex the mixing for a particular star was. 

It is worth to mention that this mixed population id is only a compact information which provides a glimpse into a history of a given star -- the higher the number, the more events the star had with different populations. However, if needed, with \MOCCA output files one can recover the full history for any given star in the system (all stellar evolutionary events, all dynamical interactions, exchanges, disruptions, collisions etc.).

The main output file which gives information about global properties of a star cluster is called \textit{system}. The file contains over 500 columns with global parameters of the star cluster computed on every time-step (e.g. total mass, total potential, and kinetic energies, number of MSs, WDs, BHs and many more).
The next important files concern events which can occur for any given binary or a single star (i.e. stellar evolution step, binary-single, binary-binary interactions, binary formation, physical collision between two single stars). 
All such events are carefully saved with all possible properties (stars' parameters before and after the interactions) which might be useful later: masses, radii, luminosities, semi-major axes, eccentricities, impact parameters, velocities, positions in the cluster and more. The next \MOCCA files contain information about binary and single stars which escaped from the cluster, parameters during any kind of kick which a star could get (e.g. supernova kick). There are Lagrangian radii computed for all populations for every time-steps, a number of different spatial profiles (e.g. surface brightness profile), and finally snapshot files which contain a number of parameters for every star and binary in a system.

The list of \MOCCA output files changes between versions. The complete and always up to date list of files, columns and their descriptions one can find on the \MOCCA web page\footnote{\url{https://moccacode.net/output/}}.

\section{BEANS script}
\label{s:AppendixB}

In appendix \ref{s:AppendixB} we present one of the scripts which was used to analyze \MOCCA simulations within \textsc{beans} software for the purpose of this paper. The goal of describing this script here is to show how one can perform a complex data analysis for large \MOCCA datasets with relatively easy to understand Apache Pig scripts. The following script produces data with number of stars or binaries from different populations measured in different ways (e.g. all binaries from the populations, only observational binaries).

The script is written in a language Apache Pig\footnote{\url{https://pig.apache.org}}, which is high level language for Apache Hadoop\footnote{\url{https://hadoop.apache.org}} platform used for distributed data analysis of Big Data. The scripts is described here line by line. Every line of interest is denoted with a comment line and a number (lines starting with characters \textit{-~-}). In the script there are places with characters \textit{[...]} which only purpose is to shorten the script and to make it easier to read. These places consist of lines which are similar to the lines which are around characters \textit{[...]} (e.g. list of columns are shorten with \textit{[...]}). The keywords of Apache Pig like \textit{load}, \textit{using}, are case insensitive.

\begin{verbatim}
-- <1>
snap = LOAD ' DATASETS="MOCCA" TABLES="snapshot"   
          FILTER="(timenr == 0) 
          	OR (tphys>400.0 tphys<430.0) 
          	[...]
          	OR (tphys>12000.0 tphys<12030.0) " ' 
        USING BeansTable();

-- <2>
sys = LOAD ' DATASETS="MOCCA" TABLES="system" ' 
        USING BeansTable();

-- <3>
aux = LOAD ' DATASETS="Auxiliary MOCCA tables" 
        TABLES="Additional lagrangian radii 
                (survey2, projection, 
                selected snapshots)" ' 
        USING BeansTable();
       
-- <4>
aux = FOREACH aux GENERATE *, 
        DSID(tbid1) as dsid; 

-- <5>
snap = FOREACH snap GENERATE tbid as tbidSnap, 
         DSID(tbid) as dsidSnap, timenr, 
         tphys, r, ik1, ik2, sm1, sm2, 
         popId1, popId2, idd1;

-- <6>
sys = FOREACH sys GENERATE tbid as tbidSys, 
        DSID(tbid) as dsidSys, 
        timenr, r_h, rhob, rtid, sturn, sturnm,
        pop1oc, pop2oc, pop1b, pop2b;
        
-- <7>
sys0 = FILTER sys BY timenr == 0;
-- <8>
sys = JOIN sys BY dsidSys, sys0 BY dsidSys;
-- <9>
sys = FOREACH sys GENERATE 
        sys::tbidSys as tbidSys, 
		sys::dsidSys as dsidSys, 
        sys::timenr  as timenr, 
        sys::r_h     as r_h, 
        [...]
        sys0::pop1b  as pop1b,
        sys0::pop2b  as pop2b;

-- <10>
snapSys = JOIN snap BY (dsidSnap, timenr), 
            sys BY (dsidSys, timenr);

-- <11>
snap = FOREACH snapSys GENERATE 
        snap::tbidSnap as tbidSnap,
        snap::dsidSnap as dsid,
        snap::timenr as timenr,
        snap::tphys as tphys,
        snap::r as r,
        snap::idd1 as idd1,
        (snap::popId1 == 1 ? 1 : 0) AS pop11, 
        (snap::popId1 == 1 
          AND snap::ik1 <= 1 ? 1 : 0) AS pop11ms, 
        [...]
        sys::pop1b  as pop1b0,
        sys::pop2b  as pop2b0;

-- <12>
snapJoined = JOIN snap BY (dsid, timenr), 
               aux BY (dsid, timenr);

-- <13>
snap = FOREACH snapJoined GENERATE 
        snap::tbidSnap AS tbidSnap,
        snap::timenr AS timenr,
        snap::tphys AS tphys,
        snap::idd1 AS idd1,
        [...]
        snap::pop1b0 as pop1b0,
        snap::pop2b0 as pop2b0,
        (CASE 
        WHEN snap::r < aux::lagrLumR1 THEN 1
        WHEN snap::r < aux::lagrLumR10 THEN 10
        [...]
        WHEN snap::r < aux::lagrLumR90 THEN 90
        ELSE 100 END) AS lumLagr,
        (CASE 
        WHEN snap::r < aux::lagrLumR1 THEN aux::lagrLumR1
        WHEN snap::r < aux::lagrLumR10 THEN aux::lagrLumR10
        [...]
        WHEN snap::r < aux::lagrLumR90 THEN aux::lagrLumR90
        ELSE 10000.0 END) AS lumLagrR;

-- <14>
snapGr = GROUP snap BY (tbidSnap, timenr, lumLagr);

-- <15>
snapFl = FOREACH snapGr GENERATE 
            SUM(snap.pop11) + SUM(snap.pop21) as pop1c, 
            SUM(snap.pop11ms) + SUM(snap.pop21ms) as pop1cms, 
            SUM(snap.pop12) + SUM(snap.pop22) as pop2c, 
            SUM(snap.pop12ms) + SUM(snap.pop22ms) as pop2cms, 
            [...]
            SUM(snap.sinpop2ms) as sinpop2cms, 
            MIN(snap.idd1) as minId, 
            FLATTEN(snap);
            
-- <16>
snapFl = FILTER snapFl BY snap::idd1 == minId;

-- <17>
snapFl = FOREACH snapFl GENERATE pop1c, pop1cms, pop2c, 
            [...]
            snap::tbidSnap as tbidSnap, 
            DSID(snap::tbidSnap) as dsid, 
            snap::timenr as timenr, 
            FLOOR(snap::tphys, 100.0) as tphys, 
            snap::idd1 as idd1,
            (snap::lumLagrR < 9999.0 ? 
              snap::lumLagrR : snap::rtid) as r,
            [...]
            snap::pop1b0   as pop1b0,
            snap::pop2b0   as pop2b0;

-- <18>
STORE snapFl INTO 'NAME="set01 obs projection"' 
   USING BeansTable();
\end{verbatim}

In principle the commands \textit{<1> - <6>} read data from different tables and store them in so called \textit{relations} for later use.

The first command (\textit{<1>}) reads data from \textsc{beans} software from all datasets with the name \textit{mocca}, from all tables with the name \textit{snapshot}. The snapshot tables in \MOCCA simulations contains the list of all stars and binaries in the system for a number of timesteps (every 50 Myrs). This instruction uses search engine and looks for every table which fulfils these query parameters. Because \SURVEYTWO contains around 250 \MOCCA simulations, this command will read this number of snapshot files. The term \textit{USING BeansTable()} means that the script should use custom reader called BeansTable. This is a custom reader, part of \textsc{beans} software, which can read multiple tables in parallel and speed up the entire script significantly. The data read from all \MOCCA snapshot files will be ``stored'' under the relation \textit{snap} (equivalent to a variable in languages like C). Later, this relation will be used to perform further operations like filtering or grouping. Apache Pig internally uses a series of low-level MapReduce jobs, which means that the data will be processed as a stream of data, in parallel. There is no need for Apache Pig to read entire tables to memory. In the command \textit{<1>} there is also a \textit{FILTER} subcommand. This is used internally by \textsc{mocca-beans} plugin and allows to speed reading data by retrieving from \MOCCA output only the rows which fulfil this \textit{FILTER} query.

The second command (\textit{<2>}) uses also the \textit{BeansTable} reader, but in this case the data are read from \textit{system} tables from \MOCCA simulations. The \textit{system} tables in \MOCCA simulations contain several hundreds of global parameters of star clusters like total mass, core radius, half-mass radius, total number of binary-single, or binary-binary interactions, total mass of WDs and many more. A few columns (e.g. turn-off mass) from these tables will be needed later to compute e.g. the observational number of binaries. Under the relation name \textit{sys} there are stored all rows from all \textit{system} tables.

The third command (\textit{<3>}) reads one table (identified by a query \textit{Additional lagrangian radii \ldots}) from a different notebook identified by a query \textit{Auxiliary MOCCA tables}. This is the way how one can merge within one \textsc{beans} script tables coming from different notebook and compute some new values. The notebook \textit{Auxuliary MOCCA tables} contains some auxiliary tables computed based on \MOCCA simulations which might be useful for any projects. One of the auxiliary tables are Lagrangian radii computed based on projected positions and luminosities of stars -- this is a table which is being read in the command \textit{<3>}. This auxiliary table is stored in the relation \textit{aux}.

The next command (\textit{<4>}) goes over every line from relation \textit{aux}, rewrites all columns (statement \textit{*}), and adds additional column (\textit{dsid}). The column is added using a function \textit{DSID}. Every table within \textsc{beans} is described by a unique ID, called hereafter \textit{TableId}. The subcommand \textit{DSID(tbid1)} takes a column \textit{tbid1} from relation \textit{aux}, returns what is the ID of a dataset for this table and saves this value under a column name \textit{dsid}. This \textit{dsid} column will be needed later to join different tables coming from different \MOCCA simulations together. 

The next command (\textit{<5>}) goes over every row from relation \textit{snap} and extracts only a few columns from it (TableId and saved it as \textit{tbidSnap}, time number \textit{timenr}, physical time \textit{tphys}, position in the cluster \textit{r}, stellar types of stars \textit{ik1}, \textit{ik2}, masses \textit{sm1}, \textit{sm2}, population ids \textit{popId1}, \textit{popId2} and id of the first star \textit{idd1}). There is also new column added here, just like in the previous command, \textit{dsidSnap}. This is ID of a dataset for every snapshot line (one \MOCCA simulation is stored in one dataset, so their IDs differ). This column will be used later to join the rows in relation \textit{snap} with the rows of other relations. 

The next command (\textit{<6>}) is very similar to the row \textit{<5>} but in this command some number of columns are extracted from \textit{system} tables (e.g. half-mass radius \textit{r\_h}, observational half-mass radius \textit{rhob}, tidal radius \textit{rtid}, turn-off mass \textit{sturn}, total number of objects from FG \textit{pop1oc}, SG \textit{pop2oc} and total number of binaries from FG \textit{pop1b}, and SG \textit{pop2b}.

The commands \textit{<7> - <8>} append to every line from \MOCCA \textit{system} tables (relation \textit{sys}) a few columns from the same tables but from physical time 0~Myr.

The next command (\textit{<7>}) read all rows from relation \textit{sys} (table \textit{system} from \MOCCA simulations) and leaves only these lines which come from the physical time 0~Myr (timenr == 0, timenr is an integer number which enumerates timesteps in \MOCCA simulations). The rows, coming from time 0~Myr, are stored separately in relation \textit{sys0}.

The next command (\textit{<8>}) uses \textit{JOIN} statement to join rows from two relations \textit{sys} and \textit{sys0}. Join operator works in such a way that for all rows from the first relation (\textit{sys}) it looks for a corresponding row from the second relation (\textit{sys0}). In out case the rows are matched using the column dsidSys for both relations. This is the column which stored IDs of datasets which particular tables belong to. In other words, with the command \textit{<8>} we add to all rows from \textit{system} tables (\textit{sys}) a few values from the same table but from time T=0~Myr (relation \textit{sys0}).

The command \textit{<9>} only rewrites the columns names to a simpler names -- Apache Pig during the last join command added prefixed like \textit{sys::} and \textit{sys0::} because there were the same columns in both relations.

In the command \textit{<10>} there is also a join operator applied, but this time for \textit{snapshot} rows (relations \textit{snap}) and rows with global parameters of star clusters (relation \textit{sys}). The purpose of this join is also to attach to snapshot rows some values of star clusters. The output of this join is stored in the relation \textit{snapSys}. The command \textit{<11>} takes as an input relation from the previous command (\textit{FOREACH snapSys}) and for every row from that relation it computes some additional values. The purpose of the entire script is to compute the number of stars, binaries for Lagrangian radii taking into account e.g. different definitions of binaries. Thus, in the command \textit{<11>} one can see different statements computing some values. For example the expression \textit{(snap::popId1 == 1 AND snap::ik1 <= 1 ? 1 : 0) AS pop11ms} checks if a first star in the \textit{snapshot} row belongs to FG and is MS star ($\rm ik1$ is stellar type in SSE/BSE code). If a star is indeed MS star from FG the number 1 is returned. There is a number of similar, quite self-explanatory expressions, like this in the command \textit{<11>}. All of these values will be needed later to compute occurrences of such stars or binaries at different Lagrangian radii. Please, notice that at the end of command \textit{<11>} there are rewritten some values from relation \textit{sys}, like half-mass radius ($\rm R_h$), observational half-mass radius (rhob). These values were attached to snapshot rows in the previous command using join operator. 

The result of the long command \textit{<11>} is stored in the relation \textit{snapJoined} and it is used in command \textit{<12>} for another join operator. This time we join the snapshot rows with Auxiliary MOCCA table which holds the Lagrangian radii for the entire system (see command \textit{<3>}). Here, the interesting part is that the join is performed for two relations \textit{snap} and \textit{aux} but with two columns: first by \textit{dsid} and then by \textit{timenr}. This, join has to be done over two columns because first one has to divide all rows by the dataset ID, and then all rows from one dataset have to be divided into separate timesteps. 

The next command \textit{13} only rewrites some columns by simplifying the names from e.g. \textit{snap::pop11} to \textit{pop11}. There are only two additional statements in this command at the end which checks for every object from snapshot rows to which Lagrangian radii the star or binary belongs (statement \textit{CASE \ldots WHEN \ldots ELSE}). The statement saves the Lagrangian radii as an integer number (1, 10, 20\ldots) and as a real position in a star clusters. 

The commands \textit{<14> - 18} in principle perform the last task of the script -- they calculate how many objects from different populations are in different Lagrangian radii. 

The command \textit{<14>} performs a group operation. This time the grouping is done over three columns: snapshot table ID (tbidSnap), timestep (timenr) and at last over the Lagrangian radius computed by the luminosity (lumLagr). In this way all rows from snapshot files  coming from all \MOCCA simulations are divided into smaller groups. In command \textit{<15>} every group is aggregated. For every Lagrangian radius it is computed e.g. how many individual stars belong to FG (\textit{SUM(snap.pop11) + SUM(snap.pop21)}), how many MS individual stars belong to FG (\textit{SUM(snap.pop11ms) + SUM(snap.pop21ms) as pop1cms}), the same for SG, then how many binaries, observational binaries, observational MS binaries belong to FG and SG etc. 

The command \textit{<16>} is used to filter many snapshot lines produced by the previous rows and leave only one row per one dataset, per one timestep and per one Lagrangian radii. The command \textit{<17>} only simplifies the columns names, computes a few new columns (dataset ID as \textit{dsid}, simplifies physical time to multiplications of the value 100.0) and replaces the last Lagrangian radius by the tidal radius (\textit{snap::lumLagrR < 9999.0 ? snap::lumLagrR : snap::rtid}).

This example \textsc{beans} script might be a little overwhelming at first, especially for the readers not familiar with such way of data analysis, but Apache Pig is considered as a scripting language with a steep learning curve. After spending some time on writing scripts in Apache Pig, everything is much easier to understand, and the possibilities of a smooth and fast data analysis of huge datasets rewards every time spend on learning this tool. The description was kept short and if a reader wishes to find more details about Apache Pig language, one is encouraged to read official documentations\footnote{\url{http://pig.apache.org/docs/r0.17.0/}} or tutorials on the \textsc{beans} web page\footnote{\url{https://beanscode.net/tutorials/}}. \textsc{beans} is the software written in a general form, and can be used to analyse any tabular datasets. For flat, plan text files it works out-of-the box, for more complex datasets, it might be easier to write a plugin for \textsc{beans}. The most important note is that \textsc{beans} can work as a central repository for a huge number of datasets coming from any sources with the ability to analyse them together in one script.



\bsp	
\label{lastpage}
\end{document}